\documentstyle{amsppt} \magnification=\magstep1 \hsize=6
truein \hcorrection{.375in} \vsize=8.5 truein
\parindent=20pt \baselineskip=14pt \TagsOnRight
\NoBlackBoxes\footline{\hss\tenrm\folio\hss}

\centerline{An Invariant Measure for the Loop Space of a }
\centerline{Simply Connected Compact Symmetric Space}

\vskip.5truein

\centerline{Doug Pickrell}
\centerline{Mathematics Department}
\centerline{University of Arizona}
\centerline{Tucson, Arizona 85721 }
\centerline{pickrell\@math.arizona.edu}

\vskip.5truein

\flushpar Abstract.  Let $X$ denote a simply connected
compact Riemannian symmetric space, $U$ the universal
covering of the identity component of the group of
automorphisms of $X$, and $LU$ the loop group of $U$.  In
this paper we prove the existence (and conjecture the
uniqueness) of an $LU$-invariant probability measure on a
distributional completion of the loop space of $X$.

\bigskip

\centerline{\S 0. Introduction.}

\bigskip

Let $K$ denote a simply connected compact Lie group, and
let $G$ denote the complexification of $K$.  In [Pi1] we
proved the following

\proclaim{Theorem (0.1)}There exists a $L_{pol}K$-biinvariant
probability measure $\mu$ on the formal completion,
$L_{formal}G$, of the complex loop group $LG$.
\endproclaim

We also conjectured that there is a unique such
biinvariant measure.

One purpose of this paper is to present a more
transparent proof of this theorem, especially in the
simplest case, $K=SU(2)$ (see \S 3).  A second purpose is
to generalize the theorem to a context in which $K$ is
replaced by a simply connected compact symmetric space
$X$.  Before stating this generalization, we will clarify
the meaning of various terms.

First $L_{pol}K$ and $L_{an}K$ denote the groups consisting of
maps from $S^1$ into $K$ which are polynomial (i.e.  have
finite Fourier series) and real analytic, respectively,
with pointwise multiplication.  The complexified loop group,
$L_{an}G=H^0(S^1,G)$, is a complex Lie group.  A neighborhood
of the identity consists of those loops which have a
unique (triangular or Birkhoff or Riemann-Hilbert)
factorization
$$g=g_{-}\cdot g_0\cdot g_{+},\tag 0.2$$
where $g_{-}\in H^0(D^{*},\infty ;G,1)$, $g_0\in G$, $g_{+}\in H^
0(D,0;G,1)$, and $D$
and $D^{*}$ denote the $\underline {closed}$ disks centered at $0$ and $
\infty$,
respectively (thus $(0.2)$ is an equality of holomorphic
functions which holds on a thin collar of $S^1$, the collar
depending upon $g$).  A model for this neighborhood is
$$H^1(D^{*},\frak g)\times G\times H^1(D,\frak g),\tag 0.3$$
where the linear coordinates are determined by
$\theta_{+}=g_{+}^{-1}(\partial g_{+})$, $\theta_{-}=(\partial g_{
-})g_{-}^{-1}$.  The (left or right) translates
of this neighborhood by elements of $L_{pol}K$ cover
$H^0(S^1,G)$.

The hyperfunction completion, $L_{hyp}G$, is modelled on the
space
$$H^1(\Delta^{*},\frak g)\times G\times H^1(\Delta ,\frak g),\tag 0.4$$
where $\Delta$ and $\Delta^{*}$ denote the $\underline {open}$ disks centered at $
0$ and
$\infty$, respectively, and the transition functions are
obtained by continuously extending the transition
functions for the analytic loop space of the preceding
paragraph.  The global definition is
$$L_{hyp}G=\lim_{r\downarrow 1}H^0(\{1<\vert z\vert <r\},G)\times_{
H^0(S^1,G)}\lim_{r\uparrow 1}H^0(\{r<\vert z\vert <1\})\tag 0.5$$
From the global definition it is clear that the group
$H^0(S^1,G)$ acts naturally from both the left and right of
$L_{hyp}G$.

The formal completion is defined in a similar way,
where $H^1(\Delta ,\frak g)$ is replaced by the corresponding space of
formal power series
$$H^1_{formal}(\Delta ,\frak g)=\{\theta_{+}=(\theta_1+\theta_2z+
..)dz,\quad\theta_i\in \frak g\}\simeq\prod_1^{\infty}\frak g.\tag 0.6$$
The global definition of the formal completion is
$$\bold LG=L_{formal}G=G(\Bbb C((z^{-1})))\times_{G(\Bbb C(z))}G(
\Bbb C((z))),\tag 0.7$$
where $\Bbb C((z))$ is the field of formal Laurent series
$\sum a_nz^n$, $a_n=0$ for $n<<0$, and $G(\Bbb C(z))=L_{pol}G$.

To summarize, there is an equivariant way to form
various ``distributional'' completions of the complex loop
group, respecting its homotopy and complex structure,
$$L_{pol}G\to L_{an}G\to L_{C^0\cap W^{1/2}}G\to L_{hyp}G\to \bold L
G\tag 0.8$$
where $\bold LG$ and $L_{hyp}G$ are ``dual'' to $L_{pol}G$ and $L_{
an}G$,
respectively, and $L_{C^0\cap W^{1/2}}G$ is essentially ``self-dual''.

An aesthetic advantage of the hyperfunction completion
is that $\Cal D$, the group of analytic orientation-preserving
diffeomorphisms of $S^1$, acts equivariantly. If $\mu$ is
supported on $L_{hyp}G$, then the conjectured uniqueness
of $\mu$ implies the conjecture that $\mu$ is $\Cal D$-invariant.

Now suppose that $X$ is a simply connected compact
symmetric space with a fixed basepoint.  From this we
obtain groups as in the following diagram,
$$\matrix &&G\\
&\nearrow&&\nwarrow\\
G_{\Bbb R}&&&&U\\
&\nwarrow&&\nearrow\\
&&K\endmatrix ,\tag 0.10$$
where $U$ is a simply connected compact group acting
isometrically and transitively on $X$, $X\simeq U/K$, $G$ is the
complexification of $U$, and $G_{\Bbb R}$ acts by automorphisms of
the dual (nonunitary type) symmetric space $G_{\Bbb R}/K$.
There are equivariant totally geodesic embeddings
$$\matrix U/K&\to&U\\
\downarrow&&\downarrow\\
G/G_{\Bbb R}&\to&G\endmatrix \tag 0.11$$
of symmetric spaces.  The main point is that the
antiholomorphic involution which fixes $L(G/G_{\Bbb R})\to LG$
extends naturally to our distributional completions of $LG$.
We define the hyperfunction and formal completions of
$L(G/G_{\Bbb R})$ as the identity components of the fixed point
sets of these extensions.  We obtain equivariant
homotopy equivalences
$$L_{pol}(U/K)\to L_{pol}(G/G_{\Bbb R})\to L_{an}(G/G_{\Bbb R})$$
$$\to L_{hyp}(G/G_{\Bbb R})\to \bold L(G/G_{\Bbb R}).\tag 0.12$$

The main result is the following

\proclaim{Theorem (0.13)} There exists a $L_{pol}U$-invariant
probability measure on the formal completion, $\bold L(G/G_{\Bbb R}
)$.
\endproclaim

We conjecture that there is a unique such invariant
measure.

A general simply connected $X$ can be written uniquely as
a product of irreducible $X$ (irreducible means that $\frak u$ does
not have any $\Theta$-invariant ideals, where $\Theta$ is the
involution of $U$ which fixes $K$).  There are two types of
irreducible $X$.  In the type I case $\frak u$ is simple.  In the
type II case $\frak u=\frak k\oplus \frak k$, where $\frak k$ is simple, $
\Theta (x,y)=(y,x)$,
and $X=K$.  Thus the essential novelty of this paper
concerns the type I case.

The plan of the paper is the following.  In \S 1 and \S 2 we
introduce the basic notation used throughout the paper.
In \S 1 we review some basic facts about triangular
decompositions and symmetric spaces (readers interested
in basic cases such as $S^2$ and $SU(2)$ can probably skip
this section entirely).  A more detailed treatment of the
statements in this section will appear in [Pi4].  In \S 2 we
consider loop spaces and their completions, and prove
the minimal structural results that we will need.

In \S 3 we will give a relatively elementary proof of
Theorem $(0.1)$ in the case $K=SU(2)$.  This proof reveals
that we can a priori compute many distributions for the
measure $\mu$ in Riemann-Hilbert coordinates, $(0.4)$.  We
briefly discuss the issue of uniqueness, but this remains
unresolved.

In \S 4 we will give a corresponding elementary proof of
Theorem $(0.13)$ in the case $S^2$.

In \S 5 and \S 6 we prove Theorem $(0.13)$ in general.  The
proofs in $\S 3$-$\S 6$ depend upon a certain compactness result
for Wiener measures parameterized by temperature (or
radius of the circle); this depends heavily on ideas of
the Malliavins ([MM]).

In \S 7 we discuss a conjectural formula for the ``diagonal
distribution'', and some of its potential implications.

\bigskip

\centerline{\S 1. Symmetric Spaces and Triangular Decompositions.}

\bigskip

Throughout the remainder of this paper, $U$ will denote a
simply connected compact Lie group, $\Theta$ will denote an
involution of $U$, with fixed point set $K$, and $X$ will
denote the quotient, $U/K$.  The space $X$ has the
structure of a simply connected symmetric space of
compact type.  Conversely, given such a space $X$,
together with the choice of basepoint, there is a
symmetric pair $(U,K)$, satisfying the conditions above,
such that $X\simeq U/K$.  To pin down $U$ in terms of $X$, we
could choose $U$ to be the universal covering of the
identity component of the group of automorphisms of $X$;
but for technical reasons, we will not assume this at
the outset.

There is a unique extension of $\Theta$ to a holomorphic
automorphism of $G$, the complexification of $U$.
Corresponding to the diagram of groups in $(0.10)$, there is
a Lie algebra diagram
$$\matrix &&\frak g=\frak u\oplus i\frak u&&\\
&\nearrow&&\nwarrow&&\\
\frak g_{\Bbb R}=\frak k\oplus \frak p&&&&\frak u=\frak k\oplus i
\frak p\\
&\nwarrow&&\nearrow\\
&&\frak k\endmatrix ,\tag 1.1$$
where $\Theta$, acting on the Lie algebra level, is $+1$ on $\frak k$ and
$-1$ on $\frak p$.  We let $(\cdot )^{-*}$ denote the Cartan involution for
the pair $(G,U)$ (so that $(\cdot )^{*}$ is an antiholomorphic
antiinvolution).  The Cartan involution for the pair
$(G,G_{\Bbb R})$ is given by $\tau (g)=g^{-*\Theta}$.  Since $*$, $
\Theta$, $\tau$, and $(\cdot )^{-1}$
commute, our practice of writing $g^{\Theta}$ for $\Theta (g)$, etc,
should not cause any confusion.

There are natural maps
$$\matrix K&\to&U&\to&U/K\\
\downarrow&&\downarrow&&\downarrow\\
G_{\Bbb R}&\to&G&\to&G/G_{\Bbb R}\endmatrix .\tag 1.2$$
The vertical arrows (given by inclusion) are homotopy
equivalences; more precisely, there are diffeomorphisms
(polar or Cartan decompositions)
$$K\times \frak p\to G_{\Bbb R},\quad U\times i\frak u\to G,\quad
U\times_Ki\frak k\to G/G_{\Bbb R},\tag 1.3$$
in each case given by the formula $(g,X)\to gexp(X)$ (mod
$G_{\Bbb R}$ in the last case).  In turn there are totally geodesic
embeddings (or morphisms) of symmetric spaces
$$\matrix U/K&@>{\phi}>>&U&:&gK&\to&gg^{-\Theta}\\
\downarrow&&\downarrow\\
G/G_{\Bbb R}&@>{\phi}>>&G&:&gG_{\Bbb R}&\to&gg^{*\Theta}\endmatrix \tag 1.4$$
The map $\phi$ is equivariant, where $g\in G$ acts on $g_0\in G$ by
$g_0\to gg_0g^{*\Theta}$.  The $\phi$-images are defined by simple
algebraic equations, modulo connectedness issues,
$$\matrix \phi (U/K)=\{g\in G:g^{-1}=g^{*}=g^{\Theta}\}_0&\to&U=\{
g^{-1}=g^{*}\}\\
\downarrow&&\downarrow\\
\phi (G/G_{\Bbb R})=\{g^{*}=g^{\Theta}\}_0&\to&G\endmatrix \tag 1.5$$
where the $0$-subscript denotes the connected component
containing the identity (Note:  the maps $(1.4)$ exist for
arbitrary automorphisms $\Theta$; the simple characterization
of the images is peculiar to automorphisms of order $2$).

Fix a maximal abelian subalgebra $\frak t_0\subset \frak k$.  We then obtain
$\Theta$-stable Cartan subalgebras
$$\frak h_0=\Cal Z_{\frak g_{\Bbb R}}(\frak t_0)=\frak t_0\oplus
\frak a_0,\quad \frak t=\frak t_0\oplus i\frak a_0\tag 1.6$$
and $\frak h=\frak h_0^{\Bbb C}$, for $\frak g_{\Bbb R}$, $\frak u$, and $
\frak g$, respectively, where $\frak a_0\subset \frak p$.
We let $T_0$ and $T$ denote the maximal tori in $K$ and $U$
corresponding to $\frak t_0$ and $\frak t$, respectively.

Let $\Delta$ denote the roots for $\frak h$ acting on $\frak g$; $
\Delta\subset \frak h_{\Bbb R}^{*}$, where
$\frak h_{\Bbb R}=\frak a_0\oplus i\frak t_0$.  We choose a Weyl chamber $
C^{+}$ which is
$\Theta$-stable (to prove that $C^{+}$ exists, we must show that
$i\frak t_0$, the $+1$ eigenspace of $\Theta$ acting on $\frak h_{
\Bbb R}$, intersects the
regular part of $\frak h_{\Bbb R}$; since $\frak t_0$ is maximal abelian, we can
find regular elements in $\frak t_0$).  Since $\tau =-(\cdot )^{*
\Theta}$ and $(\cdot )^{*}$ is
the identity on $\frak h_{\Bbb R}$, $\tau (C^{+})=-C^{+}$.

Given our choice of $C^{+}$, we obtain a $\Theta$-stable triangular
decomposition $\frak g=\frak n^{-}\oplus \frak h\oplus \frak n^{+}$, so that $
\tau (\frak n^{\pm})=\frak n^{\mp}$.  Let
$N^{\pm}=exp(\frak n^{\pm})$, $H=exp(\frak h)$, and $B^{\pm}=HN^{
\pm}$.  We also let $W$
denote the Weyl group, $W=N_U(T)/T\simeq N_G(H)/H$.

At the group level we have the Birkhoff or triangular or
LDU decomposition for $G$,
$$G=\bigsqcup_W\tilde{\Sigma}_w,\quad\tilde{\Sigma}_w=N^{-}wHN^{+}
,\tag 1.7$$
where $\tilde{\Sigma}_w$ is diffeomorphic to $(N^{-}\cap wN^{-}w^{
-1})\times H\times N^{+}$.
The intersection of this decomposition with the spaces
in $(1.5)$ is considered in detail in [Pi4].  Here we will
only prove what we need for $\phi (G/G_{\Bbb R})$.

\proclaim{ (1.8) Proposition} Fix $w\in W$.

(a) The intersection $\{g^{*}=g^{\Theta}\}\cap\tilde{\Sigma}_w$ is nonempty if and
only if there exists $\bold w\in w\subset N_U(T)$ such that $\bold w^{
*\Theta}=\bold w$;
$\bold w$ is unique modulo the action
$$T\times \{\bold w\in N_U(T):\bold w^{*\Theta}=\bold w\}\to \{\bold w^{
*\Theta}=\bold w\}:\lambda ,\bold w\to\lambda \bold w\lambda^{*\Theta}
.$$

(b) The orbits of $B^{-}$ in $\{g^{*}=g^{\Theta}\}\cap\tilde{\Sigma}_
w$ are open and
indexed by
$$\pi_0(\{\bold w\in w:\bold w^{*\Theta}=\bold w\})\simeq \{\bold w
\in w:\bold w^{*\Theta}=\bold w\}/T.$$

(c) The $B^{-}$-orbit through $\bold w$ belongs to $\phi (G/G_{\Bbb R}
)$ if and
only if $Ad(\bold w)\circ\Theta$ is equivalent to $\Theta$ through conjugation
by $Ad(K)$.

(d) For the top stratum, the map
$$N^{-}\times (T_0^{(2)}\times_{T_0^{(2)}\cap exp(i\frak a_0)}exp
(i\frak a_0))\times exp(i\frak t_0)\to \{g^{*}=g^{\Theta}\}\cap\tilde{
\Sigma}_1$$
$$l,[\bold w,m],a_{\phi}\to g=l\bold wma_{\phi}l^{*\Theta}$$
is a diffeomorphism.

\endproclaim

In this paper we will only use (d).  To prove this
suppose that $g\in\tilde{\Sigma}_1$.  There is a unique decomposition
$g=lhu$, where $l\in N^{-}$, $h\in H$, and $u\in N^{+}$.  If $g=g^{
*\Theta}$,
then $lhu=u^{*\Theta}h^{*\Theta}l^{*\Theta}$, and because $(\cdot
)^{*\Theta}$ interchanges
$N^{\pm}$, uniqueness of the decomposition implies $u=l^{*\Theta}$ and
$h=h^{*\Theta}$.  This leads to (d) in a routine way.

\proclaim{Proposition (1.10)} For the action of $U$ on $G$
given by $u,g\to ugu^{-\Theta}$, for each $U$-orbit $\Cal O$, the intersection
of $\Cal O$ with $\tilde{\Sigma}_1$ is dense in $\Cal O$.
\endproclaim

\demo{Proof of (1.10)} Let $v$ denote a highest weight
vector for the highest weight representation
corresponding to the sum of the dominant integral
functionals.  Then
$$\tilde{\Sigma}_1=\{g\in G:\langle g\cdot v,v\rangle\ne 0\},\tag 1.11$$
where $\langle\cdot ,\cdot\rangle$ denotes the essentially unique $
U$-invariant
Hermitian inner product.

Fix $g_0\in G$.  We must show that there exist $g\in U$,
arbitrarily close to $1$, such that $\langle gg_0g^{-\Theta}v,v\rangle
\ne 0$.
Suppose that this is not the case.  Then
$$\langle gg_0g^{-\Theta}v,v\rangle =0,\tag 1.12$$
for all $g\in U$ sufficiently close to $1$.  Since the left hand
side of this equation is a holomorphic function of $g\in G$,
and since $U$ is a real form of $G$, $(1.12)$ holds for all
$g\in G$.  If we take $g=b\in B^{+}$, then $b^{-\Theta}\in B^{+}$, $
b^{-\Theta}\cdot v$ is a
multiple of $v$, and $b^{*}\in B^{-}$.  Hence $(1.12)$ implies
$$\langle g_0v,b^{*}v\rangle =0,\quad\forall b^{*}\in B^{-}.\tag 1.13$$
But $\{b^{*}v:b^{*}\in B^{-}\}$ spans the space of the representation.
This is a contradiction.  \qed
\enddemo

\proclaim{Corollary (1.14)} Let $U$ act on $G$ as in $(1.10)$.
Suppose that $\nu$ is a $U$-quasi-invariant measure on $G$.
Then the $\nu$-measure of the complement of $\tilde{\Sigma}_1$ is zero.
\endproclaim

\demo{Proof of (1.14)} The measure $\nu$ will have a
disintegration along the orbits of $U$, and for $a.e.$ such
orbit, the derivative $\nu_{\Cal O}$ will belong to the unique
(Lebesgue) invariant measure class.  Proposition $(1.10)$
implies that the $\nu_{\Cal O}$-measure of $\Cal O\setminus\tilde{
\Sigma}_1$ is zero.  This
implies $(1.14)$.  \qed
\enddemo

\bigskip

\centerline{\S 2. Loop Spaces and Completions.}

\smallskip

In the introduction we recalled the definitions of $\bold LG$ and
$L_{hyp}G$ (see Part III, chapter 2, of [Pi1]).  To consider
the finer properties of loop groups
and their completions, it is useful to adopt the
Kac-Moody point of view, where many structural results
carry over from the theory of simple Lie algebras.  This
point of view will be needed only occasionally (in
the proof of Lemma $(6.5)$ and Appendix B).
 Otherwise it will suffice to
adopt a more naive approach, developed here.

These completions have generalized Birkhoff
factorizations
$$\bold LG=\bigsqcup_{Hom(S^1,T)}\tilde{\Sigma}_{\lambda}^{formal}
,\quad\tilde{\Sigma}_{\lambda}^{formal}=G(\Bbb C[[z^{-1}]])\cdot\lambda
\cdot G(\Bbb C[[z]])\tag 2.1$$
where $\Bbb C[[\zeta ]]$ denotes formal power series in $\zeta$, and
$$L_{hyp}G=\bigsqcup_{Hom(S^1,T)}\tilde{\Sigma}^{hyp}_{\lambda},\quad
\tilde{\Sigma}_{\lambda}^{hyp}=H^0(\Delta^{*},G)\cdot\lambda\cdot
H^0(\Delta ,G),\tag 2.2$$
which restrict to the standard Birkhoff factorization of
$ $$LG$ (see chapter 8 of [PS]).  It is the existence of these
coherent decompositions, corresponding to different
smoothness conditions, which imply that these various
completions are all homotopy equivalent (see $(8.6.6)$ in
[PS]).

In both the formal and hyperfunction cases, the top
stratum (the piece with $\lambda =1$ above) is open and dense,
and for each point $g$ in the top stratum, there is a
unique factorization as in $(0.2)$, where in the
hyperfunction case $g_{\pm}$ are $G$-valued holomorphic functions
in the
the open disks $\Delta$ and $\Delta^{*}$, respectively, and in the formal
case $g_{\pm}$ are simply formal power series satisfying the
appropriate algebraic equations determined by $G$.
We will refer
to $g_{-},g_0,g_{+}$ ($\theta_{-},g_0,\theta_{+}$, respectively) as the
Riemann-Hilbert coordinates (linear Riemann-Hilbert
coordinates, respectively) of $g$.

Both $(\cdot )^{*}$ and $\Theta$ act on $LG$, by acting on a loop
pointwise.

\proclaim{Lemma (2.3)} Both $(\cdot )^{*}$ and $\Theta$ extend
continuously to involutions of $L_{hyp}G$ and $\bold LG$, and in
general both permute the Birkhoff strata.
\endproclaim

\demo{Proof of (2.3)} We will consider the hyperfunction
case for definiteness.  Suppose that $g\in L_{hyp}G$.  This
means that $g=[g_l,g_r]$ is an equivalence class
represented by a pair $(g_l,g_r)$, where $g_l$ $(g_r)$ is a
holomorphic map $\{1-\epsilon <\vert z\vert <1\}\to G$ ($\{1<\vert
z\vert <1+\epsilon \}\to G$,
respectively), for some $\epsilon >0$, and this pair is equivalent
to any other pair of the form $(g_lh,h^{-1}g_r)$, for some
$h\in H^0(S^1,G)$.  The extension of $(\cdot )^{*}$ is given by
$$g^{*}=[g_r(\bar {z}^{-1})^{*},g_l(\bar {z}^{-1})^{*}].\tag 2.4$$
This is well-defined and restricts to the pointwise
action of $(\cdot )^{*}$ on $H^0(S^1,G)$, simply because $z^{-1}=
\bar {z}$ on $S^1$.

From the formula $(2.4)$, it is clear that $(\cdot )^{*}$ maps the $
\lambda$
stratum to the $\lambda^{*}=\lambda^{-1}$ stratum.

For $\Theta$, the extension is given by $g^{\Theta}=[g_l^{\Theta}
,g_r^{\Theta}]$ (i.e.  we
are simply applying $\Theta$ pointwise).  Again it is clear this
is well-defined, extends the pointwise application of $\Theta$ to
$H^0(S^1,G)$, and the $\lambda$ stratum is mapped into the $\lambda^{
\Theta}$
stratum.  \qed
\enddemo

\proclaim{Definition (2.5)} $\bold L(G/G_{\Bbb R})$ and $L_{hyp}(
G/G_{\Bbb R})$ are
the identity components of the fixed point sets of $(\cdot )^{*\Theta}$
acting on $\bold LG$ and $L_{hyp}G$, respectively.  The intersections
of $\tilde{\Sigma}_1^{formal}$ and $\tilde{\Sigma}_1^{hyp}$ with $
\bold L(G/G_{\Bbb R})$ and $L_{hyp}(G/G_{\Bbb R})$,
respectively, will be referred to as the top strata.
\endproclaim

\proclaim{Proposition (2.6)} (a) The intersection
$\tilde{\Sigma}^{formal}_{\lambda}\cap \bold L(G/G_{\Bbb R})$ is nonempty if and only if $
\lambda^{*\Theta}=\lambda$ in
$Hom(S^1,T)$.

(b) The top stratum is open and dense and diffeomorphic
to
$$G/G_{\Bbb R}\times H^1_{formal}(\Delta ,\frak g).$$
The analogous statements hold in the hyperfunction case.
\endproclaim

We only need Part (b).  This follows directly from the
uniqueness of the Birkhoff factorization when $\lambda =1$.

Finally we recall some facts about automorphisms.

Suppose that the group $U$ is the universal covering of
the identity component of $Aut(X)$.  Let $\pi$ denote the
projection, so that we have an exact sequence
$$0\to ker(\pi )\to U@>{\pi}>>Aut(X)@>{\pi_0}>>\pi_0(Aut(X))\to 0
.\tag 2.7$$
Let $A_0$ denote the identity component of $Aut(X)$.  Define
$$\Cal LA_0=\{g:\Bbb R\to U:\exists\Delta\in ker(\pi )\quad with\quad
g(t+1)=g(t)\Delta ,\forall t\},\tag 2.8$$
where some degree of smoothness ($>1/2)$) is implicitly
fixed.  There is an exact sequence
$$0\to ker(\pi )\to \Cal LA_0\to LAut(X)\to\pi_0(Aut(X))\to 0,\tag 2.9$$
where $g\in \Cal LA_0$ maps to the loop $\pi (g(e^{2\pi it}))$.  We will
indicate the degree of smoothness by attaching a
subscript, e.g.  $\Cal L_{pol}A_0$ denotes the group of elements $
g$
such that $\pi (g(e^{2\pi it}))$ is polynomial, i.e.  has a finite
Fourier series relative to some matrix representation of
$Aut(X)$.

It is straightforward to check that $\Cal L_{pol}A_0$ acts naturally
on $\bold L(G/G_{\Bbb R})$ (these actions will be written out explicitly
in subsequent sections).

These generalized loops will play a role in
this paper for the following reason.  When one fixes a
triangular decomposition for $\frak g$, there is no residual $Ad(
G)$
symmetry.  However outer automorphisms occasionally
exist that respect this
decomposition.  For the loop algebra, $L\frak g$, elements of the
center $C(K)$ give rise to outer automorphisms (see
$(3.4.4)$ of [PS]) that interact in a relatively simple way
with Riemann-Hilbert factorization.  These outer
automorphisms are represented by
multivalued loops.

\bigskip

\centerline{\S 3. The Group Case, $X=K=SU(2)$.}

\bigskip

In this section we will give a relatively elementary
proof of Theorem $(0.1)$, in the case $K=SU(2)$.  We will
then discuss a possible method for computing the
measure explicitly.

\bigskip

\flushpar {\bf Existence.  }

\smallskip

Suppose that
$$g=g_{-}\cdot g_0\cdot g_{+}\tag 3.1$$
is a point in the top stratum of $\bold LSL(2,\Bbb C)$.  We write
$$g_0=\left(\matrix a_0&b_0\\
c_0&d_0\endmatrix \right),\quad\quad a_0d_0-b_0c_0=1,\tag 3.2$$
$$g_{+}=\left(\matrix a(z)&b(z)\\
c(z)&d(z)\endmatrix \right)=1+\left(\matrix a_1&b_1\\
c_1&-a_1\endmatrix \right)z+..\tag 3.3$$
and
$$g_{-}=\left(\matrix A(z)&B(z)\\
C(z)&D(z)\endmatrix \right)=1+\left(\matrix A_1&B_1\\
C_1&-A_1\endmatrix \right)z^{-1}+..\tag 3.4$$

The action of the constants $SL(2,\Bbb C)\times SL(2,\Bbb C)$ on $
\bold LSL(2,\Bbb C)$
is completely transparent in these coordinates:  for
$g_l,g_r\in SL(2,\Bbb C)\times SL(2,\Bbb C)$,
$$g_l\cdot g\cdot g_r^{-1}=[g_lg_{-}g_l^{-1}]\cdot [g_l^{}g_0g_r^{
-1}]\cdot [g_rg_{+}g_r^{-1}].\tag 3.5$$

Let $\sigma$ denote the outer automorphism of $LSL(2,\Bbb C)$ given
by
$$\sigma (\left(\matrix a&b\\
c&d\endmatrix \right)=\left(\matrix a&bz\\
cz^{-1}&d\endmatrix \right),\tag 3.6$$
i.e.  $\sigma$ is conjugation by the multivalued loop
$\left(\matrix z^{1/2}&0\\
0&z^{-1/2}\endmatrix \right)$.  We will write out the action of $
\sigma$ on
$\bold LSL(2,\Bbb C)$ in the Lemma below.

Let $\bold w_0=\left(\matrix 0&i\\
i&0\endmatrix \right)$ and
$$i_0(SL(2,\Bbb C))=\bold w_0SL(2,\Bbb C)^{\sigma}\bold w_0^{-1}=
\{\left(\matrix d&cz^{-1}\\
bz&a\endmatrix \right):ad-bc=1\}.\tag 3.7$$
A basic fact is that $L_{pol}SL(2,\Bbb C)$ is generated by $SL(2,
\Bbb C)$
$i_0(SL(2,\Bbb C))$ (see $(5.2.5)$ of [PS]; $\bold w_0$ is perhaps a
distraction in this context; it appears here to
align our notation with that in \S 5, and its significance is
somewhat illuminated by Appendix B).

\proclaim{ (3.8) Lemma} Suppose that
$h=\left(\matrix d&cz^{-1}\\
bz&a\endmatrix \right)\in i_0(SL(2,\Bbb C))$, and $g\in \bold LG$ is in the top
stratum, as in $(3.1)$-$(3.4)$.

a) If $\alpha =a+bB_1\ne 0$, then $h\cdot g$ is also in the top stratum
and has factorization
$$[\left(\matrix d&cz^{-1}\\
bz&a\endmatrix \right)g_{-}\left(\matrix \alpha&0\\
\gamma (z)&\alpha^{-1}\endmatrix \right)]\cdot [\left(\matrix \alpha^{
-1}&0\\
-\gamma_0&\alpha\endmatrix \right)g_0]\cdot [g_0^{-1}\left(\matrix
1&0\\
\frac b{\alpha}z&1\endmatrix \right)g_0g_{+}],$$
where $\gamma_0=\frac {-2abA_1+b^2(B_2-A_1B_1)}{\alpha}$ and $\gamma
(z)=\gamma_0-bz$.

b) If $\alpha'=d-cC_1\ne 0$, then $g\cdot h^{-1}$ is in the top stratum
and has factorization
$$[g_{-}g_0\left(\matrix 1&-\frac c{\alpha'}z^{-1}\\
0&1\endmatrix \right)g_0^{-1}]\cdot [g_0\left(\matrix \alpha^{\prime
-1}&-\beta_0\\
0&\alpha'\endmatrix \right)]\cdot [\left(\matrix \alpha'&\beta (z^{
-1})\\
0&\alpha^{\prime -1}\endmatrix \right)g_{+}\left(\matrix d&cz^{-1}\\
bz&a\endmatrix \right)^{-1}]$$
where $\beta_0=\frac {2cda_1+c^2(c_2-a_1c_1)}{\alpha'}$ and $\beta
(z^{-1})=\beta_0+cz^{-1}$.

c) If $a_0\ne 0$, then $g^{\sigma}$ is also in the top stratum, and
$$(g^{\sigma})_{-}=(g_{-})^{\sigma}\left(\matrix 1&-B_1\\
\frac {c_0}{a_0}z^{-1}&1-\frac {B_1c_0}{a_0}z^{-1}\endmatrix \right
),$$
$$(g^{\sigma})_0=\left(\matrix a_0+\frac {B_1c_1}{a_0}&\frac {B_1}{
a_0}\\
\frac {c_1}{a_0}&\frac 1{a_0}\endmatrix \right),\quad (g^{\sigma}
)_{+}=\left(\matrix 1&\frac {b_0}{a_0}z\\
-c_1&1-\frac {c_1b_0}{a_0}z&\endmatrix \right)(g_{+})^{\sigma}$$
\endproclaim

\demo{Proof of (3.8)} These are straightforward matrix
calculations (they also follow from the more abstract
calculations in \S 5 below).//
\enddemo

Part a) implies
$$B((h\cdot g)_{-})=(dB+cDz^{-1})/\alpha\tag 3.9$$
$$D((h\cdot g)_{-})=(bBz+aD)/\alpha\tag 3.10$$
or in terms of components
$$B_n((h\cdot g)_{-})=(dB_n+cD_{n-1})/\alpha\tag 3.11$$
$$D_{n-1}((h\cdot g)_{-})=(bB_n+aD_{n-1})/\alpha\tag 3.12$$

This leads to the following

\proclaim{Corollary (3.13)}The space of variables
$\{B_1,D_{n-1},B_n\}$ is invariant under the action of $h$.  Define
$B_n'=B_n/D_{n-1}$.  Then
$$B_n'(h\cdot g)=\frac {c+dB_n'}{a+bB_n'},$$
i.e.  $B_n'$ is equivariant with respect to the action of
$i_0(SL(2,\Bbb C))$ on $\bold LG$ and the linear fractional action of
$SL(2,\Bbb C)$ on $\hat {\Bbb C}$.
\endproclaim

\demo{Proof of Theorem (0.1) in the case $K=SU(2)$} Let
$\nu_{\beta}$ denote Wiener probability measure for $L_{C^0}K$ with
inverse temperature $\beta$ (see Appendix A).  Via the
inclusion $L_{C^0}K\to \bold LG$, we can view the $\nu_{\beta}$ as probability
measures on the formal completion, and these measures
are quasi-invariant with respect to the left and right
action of $L_{pol}K$, and the action of $\sigma$.  This family of
measures is asymptotically invariant with respect to
these actions, meaning for example that
$$\int\vert 1-\frac {d\nu_{\beta}(g^{\sigma})}{d\nu_{\beta}(g)}\vert
d\nu_{\beta}(g)\to 0\tag 3.14$$
as $\beta\to 0$, with similar results for $g^{\sigma}$ replaced by $
l\cdot g$ or
$g\cdot r$, for $l,r\in L_{pol}K$ (see Appendix A).

\proclaim{Lemma (3.15)} The top stratum of $L_{pol}G$ has
full measure with respect to $\nu_{\beta}$.
\endproclaim

This is true for any $L_{pol}K$-biinvariant measure; see $(2.1)$
of Part I of [Pi1].  We will reproduce this argument
below when we consider symmetric spaces (see Lemma
$(6.5)$).

\proclaim{Lemma (3.16)} The measures $\nu_{\beta}$ have weak
limits, as $\beta\to 0$, with respect to $BC=BC(\tilde{\Sigma}_1^{
formal})$,
bounded continuous functions in the linear
Riemann-Hilbert coordinate space
$$H^1_{formal}(\Delta^{*},\frak g)\times G\times H^1_{formal}(\Delta
,\frak g).$$
\endproclaim

\demo{Proof of (3.16)} We will write
$$\theta_{-}=(\theta_1+\theta_2w+..)dw,\tag 3.17$$
where $w=-1/z$, and the coordinates $\theta_i\in \frak g$.  The formula
$\theta_{-}=(\partial g_{-})g_{-}^{-1}$ is equivalent to relations of the form
$$g_1=\theta_1\tag 3.18$$
$$2g_2=\theta_2+\theta_1g_1=\theta_2+\theta_1^2=\theta_2+det(\theta_
1)\tag 3.19$$
$$3g_3=\theta_3+\theta_2g_1+\theta_1g_2=\theta_3+\theta_2\theta_1
+\frac 12\theta_1(\theta_2+det\theta_1),\tag 3.20$$
i.e.  $ng_n=\theta_n$ plus a polynomial in lower order terms.

Slightly abusing standard terminology, we will say that a
coordinate, or more generally a function of such
coordinates, say $\lambda$, is tight if the mass of the measures
$\lambda_{*}\nu_{\beta}$ does not escape to infinity, i.e.  given $
\epsilon >0$, there
is a compact set $K_{\epsilon}$ in the target space of $\lambda$ such that
$\lambda_{*}\nu_{\beta}(K_{\epsilon})>1-\epsilon$ for all $\beta$.  Sums and products of such
tight variables are also tight.  By a theorem of Prohorov
(see chapter 2, \S 6, of [Bi]), to show that $\nu_{\beta}$ has weak
limits with respect to $BC(\tilde{\Sigma}^{formal}_1)$, it is necessary and
sufficient to show that $g_0$, each of the individual
coordinates $\theta_n$, and the corresponding coordinates for $\theta_{
+}$,
are tight.

Since the functions $B_n'$ are equivariant with respect to
the action of $i_0(SU(2))$, it follows from asymptotic
invariance that
$$\lim_{\beta\to 0}(B_n')_{*}\nu_{\beta}=\frac 1{\Cal Z}(1+\vert
B_n'\vert^2)^{-2}dm(B_n'),\tag 3.21$$
the unique invariant probability for unitary linear
fractional transformations.  In particular each of the $B_n'$
is tight.

To show that each $\theta_n$ is tight, we argue as follows (this
is the same argument as in [Pi1], but it is more explicit).
We know that $B_1$ is tight (since we know the
distribution of the limit, by $(3.21)$).  The invariance with
respect to $SU(2)$ implies that $g_1=\theta_1$ is tight (From an
abstract perspective, $B_1$ is a linear function on the $\{\theta_
1\}$;
$SU(2)$ acts irreducibly by the adjoint action on $\{\theta_1\}$,
hence the $SU(2)$ orbit of $B_1$, which consists of linear
functions all having the same distributional properties,
spans the dual of $\{\theta_1\}$; this implies that $\theta_1$ is tight).
This implies that $D_1$ is tight.  Since $B_2'$ is tight, this
implies that $B_2=D_1B_2'$ is tight.  Since $\theta_2=2g_2$ plus a
polynomial function of $g_1$, it follows that $\beta_2$ is tight (in
fact a special feature of the $SU(2)$ case is that
$\beta_2=2B_2$, i.e.  there are no lower order terms).  The
invariance with respect to $SU(2)$ implies that $\theta_2$ is
tight.  This implies that $D_2$ is tight; together with the
tightness of $B_3'$ this implies $B_3$ is tight, etc.

The tightness of the coefficients of $\theta_{+}$ is proved in a
similar way.

The new aspect of our argument is a simple way to see
that $g_0$ is tight.  Because the $(g_0)_{*}\nu_{\beta}$ are
$SU(2)$-biinvariant, it suffices to show that $a_0$ is tight.

The basic observation is that the formula for $(g^{\sigma})_0$
implies that
$$\frac {b_0}{a_0}(g^{\sigma})=\frac {(B_1/a_0)}{a_0+B_1c_1/a_0}=\frac {
B_1}{a_0^2+B_1c_1}\tag 3.22$$
is tight.  But we already know that $B_1$ and $c_1$ are tight,
and the limiting distribution of $B_1$ is in the Lebesgue
class.  Hence it follows that $a_0$ is tight.  This proves
Lemma $(3.16)$.  \qed
\enddemo

We now complete the proof of $(0.1)$ in the following way
(this part of the argument will not be repeated in
subsequent sections).  We identify $BC$ with a space of
Borel functions on $\bold LG$, by extending each such function
to be zero on the complement of $\tilde{\Sigma}_1^{formal}$.  The Lemma
implies that the family $\{\nu_{\beta}\}$ has weak limits with
respect to $BC$.  If $\nu$ is a limit point, i.e.  for some
sequence $\beta_j\to 0$, $\nu_{\beta_j}(f)\to\nu (f)$, for all $f
\in BC$, then for
$g\in L_{pol}K$, asymptotic invariance implies that
$g_{*}\nu_{\beta_j}(f)\to\nu (f)$ as well.  In other words $\nu_{
\beta_j}$ has a weak
limit with respect to $g^{*}BC$, for each $g\in L_{pol}K$.  Let $
\Cal F$
denote the $L_{pol}K$-invariant space generated by $BC$, which
we can identify with a space of Borel functions on $\bold LG$.
We then have a surjective map $\oplus g^{*}BC\to \Cal F$ and an
injective map
$$\Cal F^{*}\to\prod_{L_{pol}K}\{g^{*}BC(\tilde{\Sigma}^{formal}_
1)\}^{*}.\tag 3.23$$
The image of the family $\{\nu_{\beta}\}$ is precompact in each
factor of $(3.23)$, hence the image of the family under the
map $(3.23)$ is precompact.  Suppose that $\nu$ is a limit
point.  To check that $\nu\in \Cal F^{*}$, suppose that $\sum g_i^{
*}f_i=0$ in
$\Cal F$, where $f_i\in BC$ and $g_i\in L_{pol}K$, $i=1,..,n$.  Then
$$\nu (\sum g_i^{*}f_i)=\sum\lim_{\beta_j\to 0}\nu_{\beta_j}(g_i^{
*}f_i)=\lim_{\beta_j\to 0}\nu_{\beta_j}(0)=0.\tag 3.24$$
Thus $\nu$ is a linear functional on $\Cal F$, and we can interpret
$\nu$ as a probability measure, supported on $\tilde{\Sigma}_1^{f
ormal}$ (or any
of its $L_{pol}K$-translates.

Asymptotic invariance implies that any such limit point
will be $L_{pol}K$-invariant.  \qed
\enddemo

\flushpar Remarks $(3.25)$.  (a) There is another approach
to the tightness of the variables $\theta_n$, for $n>1$, which
uses the action of $\sigma$, rather than the auxiliary variables
$B_n'$.  As before it is first necessary to observe that $B_1$
and $c_1$ are tight.  We use the $\sigma$ action to get tightness
of $a_0$, hence of $g_0$.  We then move from degree to
degree using the action of $\sigma$.  The basic fact is the
following:
$$B(g^{\sigma})=A((g_{-})^{\sigma})(-B_1)+B((g_{-})^{\sigma})(1-\frac {
B_1c_0}{a_0}z^{-1})\tag 3.26$$
$$=-A(g_{-})B_1+zB(g_{-})(1-\frac {B_1c_0}{a_0}z^{-1}).\tag 3.27$$
This implies
$$B_n(g^{\sigma})=-A_nB_1+B_{n+1}-\frac {B_nB_1c_0}{a_0}.\tag 3.28$$
Using asymptotic invariance with respect to $\sigma$, this
formula shows that $B_{n+1}$ will be tight if we know that
the coefficients of $g_{-}$ are tight up to order $n$.

This argument works without change in the $S^2$ case, as
we will observe in $\S 4$.

(b) There is another approach to the tightness of $a_0$,
also using the existence of $\sigma$.  The formula for $g^{\sigma}_
0$
shows that $a_0$ and $1/a_0$ have the same limiting
distributional properties (a symmetry which we explore
in Appendix B).  Thus if $a_0$ is not tight, then
probabilistic mass escapes to $a_0=0$.  This leads to a
contradiction with Corollary $(1.14)$, modulo details
that will be handled in Lemma $(5.15)$.

\bigskip

\flushpar {\bf On Calculating the Invariant Measure}.

\smallskip

Suppose that $\mu$ is a $L_{pol}K$-biinvariant probability
measure on $\bold LG$.  We would like to calculate the $\theta_{-}$
distribution of $\mu$.

The proof of Lemma $(3.16)$ suggests the following
approach.  We know the $B_1$ distribution.  We use
$SU(2)$-invariance to determine the $\theta_1$ distribution.  This
determines the joint distribution of $D_1$ and $B_1$.  By $(3.13)$
we know the $B_2'$ distribution.  The invariant action $(3.13)$
of $i_0(SU(2))$ then determines the joint distribution of $D_1$, $
B_1$
and $B_2$.  In this $SU(2)$ case we have the special fact
that $2B_2=\beta_2$ (see $(3.19)$).  The knowledge of the $2B_2=\beta_
2$
distribution, together with $SU(2)$-invariance, then
determines the $\theta_2$ distribution.  We should then have
enough information to compute the joint distribution of
$\theta_1$ and $\theta_2$ (see $(3.29)$ below).  At least in principle, this
determines the joint distribution of $D_2$ and $B_1$.  Using
the action $(3.13)$ again, we determine the joint
distribution of $D_2$, $B_1$, and $B_3$.  This determines $\beta_
3$, and
we continue in this way.

This strategy depends upon several conjectural
uniqueness statements, and in practice the calculations
rapidly become prohibitive.  Here I will simply
summarize a partial result, and state a conjecture.

\proclaim{Proposition (3.29)} The probability measure
$$\Cal Z^{-1}(1+\vert\theta_1\vert^2+\frac 12\vert\theta_2\vert^2
)^{-7}dm(\theta_1,\theta_2)\tag 3.30$$
has the property that it restricts to $SU(2)$-invariant
distributions for $\theta_1$ and $\theta_2$, a $i_0(SU(2))$-invariant
distribution on $\{B_1,D_1,B_2\}$, and has $B_1$ and $B_2'$
distributions given by $(3.21)$.
\endproclaim

As stated, this is a straightforward calculation.  This
uses a special fact about $sl(2,\Bbb C)$, namely that $2B_2=\beta_
2$
(see $(3.19)$).  This proposition suggests, but does not
guarantee, that $(3.30)$ is the $\{\theta_1,\theta_2\}$-distribution of $
\mu$,
because there is not an accompanying uniqueness
statement (the $\{\theta_i,i\le 3\}$-distribution definitely has a
more complicated form).

Our conjecture that $\mu$ is unique, hence $\Cal D$-invariant,
implies the conjecture that the distributions of the
matrix coefficients of $\theta_{-}$ are $PSU(1,1)$-invariant.  The
current state of knowledge regarding $PSU(1,1)$-invariant
probability measures on $H^1(\Delta )$ (or more generally, $H^m(\Delta
)$)
is at the stage of constructing interesting examples
([Pi3]).  If this symmetry has some power (and I
emphasize that I have little feeling for whether this is
true), then $(3.29)$ suggests the following

\proclaim{Possible Conjecture (3.31)} The $\beta$-distribution of
$\mu$ is the probability measure with $\beta_1,..,\beta_N$ distribution
$$\Cal Z^{-1}(1+\vert\beta_1\vert^2+..+\frac 1N\vert\beta_N\vert^
2)^{-1-3N}dm(\beta_1,..,\beta_N)$$
for each $N$.
\endproclaim

This measure is a quotient of a Gaussian.  This suggests
that there might be some kind of ``free field'' (or Verma
module type) construction of the $\theta_{-}$ distribution of $\mu$.

\smallskip

\flushpar Remark (3.32).  The measure $\mu$ has a deformation
$\mu_l$ parameterized by ``level'' $l$ (extensively discussed, but
not completely proven to exist, in Part III of [Pi1]).
This parameter fits into the formulae above in a very
natural way.  For example in the case of $(3.31)$, one
simply replaces the exponent $1+3N$ by $1+3N+l$; the
corresponding measures are finite (and coherent)
provided $l>-1$.

\bigskip

\centerline{\S 4. The $S^2$ Case}

\bigskip

In this case $(0.10)$ specializes to
$$\matrix &&G=SL(2,\Bbb C)\\
&\nearrow&&\nwarrow\\
G_{\Bbb R}=SU(1,1)&&&&U=SU(2,\Bbb C)\\
&\nwarrow&&\nearrow\\
&&K=U(1)\endmatrix ,\tag 4.1$$
where
$$\left(\matrix a&b\\
c&d\endmatrix \right)^{\Theta}=\left(\matrix a&-b\\
-c&d\endmatrix \right).\tag 4.2$$
and $(\cdot )^{*}$ is the usual Hermitian conjugate.

The top stratum of $\bold L(SL(2,\Bbb C)/SU(1,1))$ consists of
$g\in \bold LSL(2,\Bbb C)$ as in $(3.1)-(3.4)$ with
$$g_0=\left(\matrix a_0&b_0\\
-\bar {b}_0&d_0\endmatrix \right),\quad a_0,d_0\in \Bbb R\tag 4.3$$
and $g_{+}=g_{-}^{*\Theta}$, so that $a_i=\bar {A}_i$, $b_i=-\bar {
C}_i$, $c_i=-\bar {B}_i$, $d_i=\bar {D}_i$.

The action of $\sigma$, from $\S 3$, commutes with the pointwise
action of $\Theta$ and $(\cdot )^{*}$.  Hence $\sigma$ will act on the formal
completion of $L(G/G_{\Bbb R})$.

The group $\Cal L_{pol}SO(3)$ is generated by $SU(2)$ (the
constants), and the map $\tilde{\sigma }:t\to\left(\matrix e^{i\pi
t}&0\\
0&e^{-i\pi t}\endmatrix \right)$; note that the
image of $\tilde{\sigma}$ in $L_{pol}Aut(S^2)_0=L_{pol}SO(3)$ is $
\sigma$.  This
follows from $(5.2.5)$ of [PS].

In this special case we will now write out the action of
$L_{pol}U$ on $\bold L(G/G_{\Bbb R})$ in Riemann-Hilbert coordinates.
These calculations follow directly from the group case
of the preceding section.

\proclaim{Lemma (4.4)} Suppose that
$h=\left(\matrix d&cz^{-1}\\
bz&a\endmatrix \right)\in i_0(SL(2,\Bbb C))$, and $g\in \bold L(G
/G_{\Bbb R})$ is in the
top stratum, as in $(3.1)-(3.4)$ and $(4.2)$.  If
$\vert a+bB_1\vert^2\ne\vert b\vert^2a_0^2$ and $a+bB_1\ne 0$, then $
g'=h\cdot g\cdot h^{*\theta}$ is
also in the top stratum and has factorization
$g'=g_{-}'\cdot g_0'\cdot g_{-}^{\prime *\theta}$, where
$$g_{-}'=\left(\matrix d&cz^{-1}\\
bz&a\endmatrix \right)g_{-}g_0\left(\matrix 1&\frac {-\bar {b}}{\bar {
a}+\bar {b}\bar {B}_1}z^{-1}\\
0&1\endmatrix \right)g_0^{-1}\left(\matrix \alpha&0\\
\gamma (z)&\alpha^{-1}\endmatrix \right),$$
$$g_0'=\left(\matrix \alpha^{-1}&0\\
-\gamma_0&\alpha\endmatrix \right)g_0\left(\matrix \alpha^{\prime
-1}&\beta_0\\
&\alpha'\endmatrix \right),$$
$$\alpha =a+b(B_1-\frac {\bar {b}}{\bar {a}+\bar {b}\bar {B}_1}a_
0^2)=\frac {\vert a+bB_1\vert^2-\vert b\vert^2a_0^2}{\bar {a}+\bar {
b}\bar {B}_1},$$
$\gamma_0=\frac {-2abA_1'+b^2(B'_2-A'_1B'_1)}{\alpha}$, $\gamma (
z)=\gamma_0-bz$, $\alpha'=\bar {a}+\bar {b}\bar {B}_1$, and
$\beta_0=\frac {2cda_1+c^2(c_2-a_1c_1)}{\alpha'}$.

The action of $\sigma$ is the same as in the group case.
\endproclaim

We will not use the precise formulas for $\beta_0$ and $\gamma_0$.

Let $\bold b=\bar {b}/(\bar {a}+\bar {b}\bar {B}_1)$, $g_{-}'=\left
(\matrix A'&B'\\
C'&D'\endmatrix \right)$.  Lemma $(4.4)$ implies
the formulas:
$$B(g'_{-})=[(dB+cz^{-1}D)-(dA+cz^{-1}C)\bold ba_0^2z^{-1}+(dB+cz^{
-1}D)\bold b\bar {b}_0a_0z^{-1}]/\alpha\tag 4.5$$
$$D(g'_{-})=[(bzB+aD)-(bzA+aC)\{\bold ba_0^2z^{-1}\}+(bzB+aD)\bold b
\bar {b}_0a_0z^{-1}]/\alpha\tag 4.6$$
or
$$B_n(g'_{-})=[(dB_n+cD_{n-1})-(dA_{n-1}+cC_{n-2})\bold ba_0^2+(d
B_{n-1}+cD_{n-2})\bold b\bar {b}_0a_0]/\alpha\tag 4.7$$
$$D_{n-1}(g'_{-})=[(bB_n+aD_{n-1})-(bA_{n-1}+aC_{n-2})\bold ba_0^
2+(bB_{n-1}+aD_{n-2})\bold b\bar {b}_0a_0]/\alpha\tag 4.8$$

\proclaim{Corollary (4.9)} The space of variables $\{a_0,B_1\}$
is invariant under the action of $h\in i_0(SL_2(\Bbb C))$; the action
is given by
$$a(g_0')=\frac {a_0}{\alpha\alpha'}=\frac {a_0}{\vert a+bB_1\vert^
2-\vert b\vert^2a_0^2},\tag 4.10$$
$$B_1(g_{-}')=[dB_1+c-\frac {d\bar {b}a_0^2}{\bar {a}+\bar {b}\bar {
B}_1}]/\alpha .\tag 4.11$$
The space of variables $\{a_0,B_1,D_1,B_2\}$ is invariant; the
action is given by $(4.10)$, $(4.11)$, and
$$B_2(g'_{-})=[dB_2+cD_1-(d(-D_1)+c)\bold ba_0^2+(dB_1+c)\bold b\bar {
b}_0a_0)]/\alpha\tag 4.12$$
$$D_1(g'_{-})=[bB_2+aD_1-b(-D_1)\bold ba_0^2+(bB_1+a)\bold b\bar {
b}_0a_0]/\alpha\tag 4.13$$

\endproclaim

The abstract explanation for the existence of this
equivariant projection, in the case $n=1$, is given by the
diagram $(6.10)$ in \S 6.

\demo{Proof of Theorem (0.13) for $X=S^2$}

Let $\tilde{\nu}_{\beta}$ denote the projection of the Wiener measure $
\nu_{\beta}$
on $L_{C^0}SU(2)$ to $L_{C^0}S^2$.  We view $\tilde{\nu}_{\beta}$ as a measure on
$\bold L(SL(2,\Bbb C)/SU(1,1))$.

\proclaim{Lemma (4.14)} $\tilde{\nu}_{\beta}$ has full measure on the top
stratum.
\endproclaim

This is a special case of Lemma $(6.5)$ (I do not know of
an argument that takes advantage of the small rank).

As in the group case, it suffices to show that the
coordinates $g_0$, $\theta_1$,..  are tight, relative to the family of
measures $\tilde{\nu}_{\beta}$.  We will first show that $g_0$ is tight.

Recall the Cartan isomorphism
$$U\times_Ki\frak k\to G/G_{\Bbb R}\to\phi (G/G_{\Bbb R}):[\left(\matrix
a&b\\
-\bar {b}&\bar {a}\endmatrix \right),\left(\matrix x&0\\
0&-x\endmatrix \right)]\to$$
$$\left(\matrix a_0&b_0\\
-\bar {b}_0&d_0\endmatrix \right)=\left(\matrix \vert a\vert^2e^{
2x}-\vert b\vert^2e^{-2x}&ab(e^{2x}+e^{-2x})\\
-\bar {a}\bar {b}(e^{2x}+e^{-2x})&\vert a\vert^2e^{-2x}-\vert b\vert^
2e^{2x}\endmatrix \right).\tag 4.15$$
The corresponding $U$-equivariant projection
$$\phi (G/G_{\Bbb R})\to U/K\tag 4.16$$
is given by
$$\left(\matrix a_0&b_0\\
-\bar {b}_0&d_0\endmatrix \right)\to\zeta =\frac {-\bar {b}_0}{a_
0+d_0+\sqrt {1+(\frac {a_0-d_0}2)^2}}\tag 4.17$$
$$=\frac {-\bar {b}_0a_0}{a_0^2+1-\vert b_0\vert^2+\frac 12(4a_0^
2+(a_0^2+\vert b_0\vert^2-1)^2)^{1/2}}\tag 4.18$$
where $\zeta =-\bar {b}/a$ is the usual affine coordinate for $\Bbb C
\Bbb P^1$.
To see this, note that $(4.15)$ implies
$$a_0+d_0=2(\vert a\vert^2-\vert b\vert^2)ch(2x)=2(2\vert a\vert^
2-1)ch(2x)\tag 4.19$$
$$a_0-d_0=2sh(2x),\quad b_0=2abch(2x)\tag 4.20$$
We can solve for $sh(2x)$ and $ch(2x)$ using the second
equation (note that $a_0-d_0$ will be our basic $U$-invariant
quantity; we will use this below).  We can then solve
the first equation for $\vert a\vert^2$ and the last equation for $
ab$.
We then divide to obtain the formula above.

Since $\tilde{\nu}_{\beta}$ is $SU(2)$-invariant, the $\zeta$-distribution of $
\nu_{\beta}$ is
the usual invariant measure $\Cal Z^{-1}(1+\vert\zeta\vert^2)^{-2}
dm(\zeta )$.

To show that $g_0$ is tight, it suffices to show $a_0$, as a
map into $\Bbb R$, is tight ($a_0$ and $d_0$ have the same
distribution and $a_0d_0=1-\vert b_0\vert^2$).  To accomplish this we
will use the action of $\sigma$.  An important technical point
is that the map
$$LSU(2)\to\phi (LS^2):g\to gg^{-\Theta}\tag 4.21$$
is $\sigma$ equivariant, since $\sigma$ is conjugation by a multivalued
loop which is fixed by $\Theta$ for each $t$.  This justifies our
use of asymptotic invariance results for $\sigma$ and $\nu_{\beta}$.

Using the formula for $(g^{\sigma})_0$ in part (c) of $(3.8)$ (which
relates $1/a_0$ and $d_0$), and then conjugating by $\left(\matrix
0&i\\
i&0\endmatrix \right)$, we
see that $a_0$ has the same limiting distribution properties
as $1/a_0$.  It thus suffices to show that $1/a_0$, as a map
into $\Bbb R$, is tight.

\proclaim{Lemma (4.22)} We identify $SL(2,\Bbb C)/SU(1,1)$
with its $\phi$-image.  Suppose that we are given a family
$\{\lambda_{\beta}:\beta >0\}$ of probability measures on $SL(2,\Bbb C
)/SU(1,1)$
which is quasi-invariant and asymptotically invariant
with respect to $SU(1,1)$.  Then $1/a_0$, as a map into $\Bbb R$, is
tight.
\endproclaim

Note that in our present context, the $\lambda_{\beta}=(g_0)_{*}\nu_{
\beta}$ are
$SU(2)$-invariant.

\demo{Proof of (4.22)}Note that by Corollary $(1.14)$, for
each $\beta$, $a_0$ is nonzero a.e.  $[\lambda_{\beta}]$.

By way of contradiction, suppose that $1/a_0$ is not tight.
Then there exists an $\epsilon >0$ such that for all $k$ there is
$\beta_k$ ($\to 0$) such that $\nu_{\beta_k}\{\vert a_0\vert <k^{
-1}\}<\epsilon$.  We would like to
assert that this means that positive probability
accumulates on $\{\left(\matrix 0&b_0\\
-\bar {b}_0&d_0\endmatrix \right):\vert b_0\vert^2=1\}$, in the limit $
\beta_k\to 0$,
and this positive mass will be $SU(2)$-invariant,
contradicting Corollary $(1.14)$.  However $\vert a_0\vert <M$ is not
compact, so mass could possibly escape along the
noncompact directions of the lower strata.  We will use
$\zeta$ to fill this gap.

We estimate $\zeta$ using $(4.18)$:
$$\vert\zeta\vert\le\vert a_0\vert\frac {\vert b_0\vert}{\vert\vert
b_0\vert^2-1\vert}=\vert a_0\vert\frac 1{\vert\vert b_0\vert -\vert
b_0\vert^{-1}\vert}.\tag 4.23$$
Since $\zeta$ is an affine coordinate for the $2$-sphere, and
represents a $SU(2)$-equivariant map, the $\zeta$-distribution of
$\nu_{\beta}$, in the limit $\beta\to 0$, is the invariant measure
$\Cal Z^{-1}(1+\vert\zeta\vert^2)^{-2}dm(\zeta )$.  Thus we can find $
M=M_{\epsilon}$ such that
$\nu_{\beta}\{\frac 1M\le\vert\zeta\vert \}\ge 1-\epsilon /2$, for all sufficiently small $
\beta$.  So with
$\nu_{\beta_k}$ probability $>\epsilon /2$,
$$\vert\vert b_0\vert -\frac 1{\vert b_0\vert}\vert <\frac Mk\quad
\implies\quad\vert\vert b_0\vert -1\vert <\frac Mk.\tag 4.24$$
In the same way, using the $(4.17)$, we obtain
$$\frac 1M\le\vert\zeta\vert\le\frac {\vert b_0\vert}{\vert a_0+d_
0\vert}\quad\implies\quad\vert a_0+d_0\vert\le M\vert b_0\vert .\tag 4.25$$
Thus we do obtain a nontrivial $SU(2)$-invariant measure
on the lower strata, in the limit $k\to\infty$, contradicting
$(1.14)$.  \qed
\enddemo

We now know that $g_0$ is tight.  To show that the $\theta_j$ are
tight, we can proceed as in the $SU(2)$-case, for which
we offered two arguments.  The first is to note that
because $a_0$ is tight, taking $h=\left(\matrix 0&z^{-1}\\
-z&0\endmatrix \right)$ in $(4.10)$, we see
that $B_1$ is tight.  Using $SU(2)$-invariance we obtain that
$\theta_1$ is tight.  This implies $a_0$, $B_1$, and $D_1$ are tight.  Using
$(4.13)$ with $h$ above, we see that $B_2$ is tight.  We now
continue, using $(4.8)$ in general.

The alternative argument is to observe that the $\sigma$-action
implies that $B_1$ has the same distributional properties as
$b_0/a_0$, hence that $B_1$ is tight.  The argument then
proceeds exactly as in our second argument in the $SU(2)$
case, part (a) of Remark $(3.25)$, to show that $\theta_{-}$ is tight.
\qed
\enddemo

\flushpar {\bf On Calculating the Invariant Measure}.

\smallskip

Suppose that $\nu$ is a $L_{pol}SU(2)$-invariant probability
measure on $\bold L(SL(2,\Bbb C)/SU(1,1))$.  The distributions which
we can a priori compute are $\zeta$ in $(4.17)$, and translates
of this (which are given by more complicated formulas).
This contrasts sharply with the group case, where we a
priori know the distribution of relatively simple
variables such as the $B_n'$.

In \S 7 we will discuss a conjectural formula for the $g_0$
distribution.  In terms of the Cartan and matrix
coordinates in $(4.15)$ (with $z=b_0/a_0$), this conjecture
reads
$$(g_0)_{*}\nu =\frac 1{\Cal Z}sech^3(2x)ch^2(2x)dk\times dx\tag 4.26$$
$$=\frac 1{\Cal Z}(1+(\frac {a_0-d_0}2)^2)^{-3/2}a_0da_0dm(z)\tag 4.27$$
$$=\frac 1{\Cal Z}(1+(\frac {(1+\vert z\vert^2)a_0^2-1}{2a_0})^2)^{
-3/2}d(a_0^2)dm(z)\tag 4.28$$
[The first expression is from Lemma $(7.5)$; the second is
obtained using $(4.20)$ to convert $sech^3(2x)$ into matrix
coordinates, and $(C.19)$ of Appendix C to get an
expression for the $G$-invariant measure in terms of $a_0$
and $z$].

Now the distribution of $z=b_0/a_0$ is the same as $B_1$
(using the formula for $\sigma$).  Therefore the diagonal
distribution conjecture implies the conjecture that the
$B_1$ distribution is given by
$$\frac 1{\Cal Z}\frac 1{(1+\vert B_1\vert^2)^{3/2}}F(\frac 1{1+\vert
B_1\vert^2})dm(B_1),\quad F(\rho )=\int_0^{\infty}\frac {\rho}{(\rho
+\frac {(x-1)^2}x)^{3/2}}dx\tag 4.29$$
The algebraic part of $(4.29)$ is reminiscent of $(3.31)$ with
the level $l=-1/2$ inserted (see Remark $(3.32)$ for the
meaning of level).

The function $F(\rho )$ tends to the finite limit
$\int_{-\infty}^{\infty}(1+x^2)^{3/2}dx$ as $\rho\downarrow 0$; it is a decreasing function, and
asymptotically behaves like $\rho^{-1}$ as $\rho\to\infty$ [To see this
observe that the integral outside the interval $\vert x-1\vert <\delta$
vanishes in the limit $\rho\to 0$; for $x$ close to $1$ one can get
rid of the solitary $x$ in the denominator of the integrand
of $(4.29)$].

At this point I am stuck.  I do not see how to use
$SU(2)$-invariance to generate an expression for the $\theta_1$
distribution, and in thinking about $PSU(1,1)$-invariant
measures, I have never come across something like $F$.

\bigskip

\centerline{\S 5. General Group Case}

\bigskip

The point of this section is to indicate how to generalize
the argument given in \S 3, especially to identify the
analogue of the variables $B_n'$, and the generalization of
the distributional symmetry of $a_0$ and $1/a_0$.

We assume that $\frak g$ is simple.  In this case there is a
highest root $\theta$.  Let $h_{\theta}$ denote the coroot, and choose $
e_{\pm\theta}$
in the $\pm\theta$ root spaces such that $e_{-\theta},h_{\theta},
e_{\theta}$ satisfy the
canonical $sl(2,\Bbb C)$ relations.  We define $i_0:sl(2,\Bbb C)\to
L_{pol}\frak g$
by
$$i_0(\left(\matrix 0&0\\
1&0\endmatrix \right))=e_{\theta}z^{-1},\quad i_0(\left(\matrix 1&
0\\
0&-1\endmatrix \right))=-h_{\theta},\quad i_0(\left(\matrix 0&1\\
0&0\endmatrix \right))=e_{-\theta}z.\tag 5.1$$
We denote the corresponding group homomorphism by the
same symbol, $i_0:SL(2,\Bbb C)\to L_{pol}G$.  When $G=SL(2,\Bbb C
)$, this
agrees with our notation in \S 3 and \S 4.  The group $L_{pol}G$
is generated by $G$ and the image of $i_0$, and a similar
statement applies with $K$ in place of $G$ ($(5.2.5)$ of [PS]).

\proclaim{Lemma (5.2)}Suppose that $\left(\matrix a&b\\
c&d\endmatrix \right)\in SL(2,\Bbb C)$ and
$g=g_{-}\cdot g_0\cdot g_{+}$ is a point in the top stratum of $\bold L
G$.  Let
$h=i_0(\left(\matrix a&b\\
c&d\endmatrix \right))$, write
$$g_{-}=exp(x_1z^{-1}+x_2z^{-2}+..),\tag 5.3$$
and $x_n=x_n'+Z_ne_{\theta}$ for the partial root space
decomposition of $x_n$.

(a) If $a+bZ_1\ne 0$, then the point
$$g'=h\cdot \{g_{-}\cdot g_0\cdot g_{+}\}$$
is also in the top stratum, and has triangular
factorization $g'=g_{-}'\cdot g_0'\cdot g_{+}'$, where
$$g_{-}'=hg_{-}exp(-\frac b{a+bZ_1}e_{-\theta}z)(a+bZ_1)^{h_{\theta}}
l_0^{-1},$$
$$g_0'=l_0(a+bZ_1)^{-h_{\theta}}g_0,\quad g_{+}'=g_0^{-1}exp(\frac
b{a+bZ_1}e_{-\theta}z)g_0g_{+},$$
and
$$l_0=exp(-ab[e_{-\theta},a^{-ad(h_{\theta})}(x_1')]-b^2Z_2e_{-\theta}
)\in N^{-}.$$

(b) If
$$g_{+}=exp(y_1z^{-1}+y_2z^{-2}+..),$$
$y_n=y_n'+W_ne_{-\theta}$, and $a+cW_1\ne 0$, then the point
$$g'=\{g_{-}\cdot g_0\cdot g_{+}\}\cdot h$$
is also in the top stratum, and has triangular
factorization $g'=g_{-}'\cdot g_0'\cdot g_{+}'$, where
$$g_{-}'=g_{-}g_0exp(\frac c{a+cW_1}e_{\theta}z^{-1})g_0^{-1},$$
$$g_0'=g_0(a+cW_1)^{-h_{\theta}}u_0,\quad$$
$$g_{+}'=u_0^{-1}(a+cW_1)^{h_{\theta}}exp(\frac {-c}{a+cW_1}e_{\theta}
z^{-1})g_{+}h,$$
and
$$u_0=exp(-ac[e_{\theta},a^{-ad(h_{\theta})}(y_1')]-c^2W_2e_{\theta}
)\in N^{+}.$$

\endproclaim

\demo{Proof of (5.2)}.  This is an easy calculation, once
one understands the basic idea.  In terms of Lie algbras,
$$\{x\in H^0(D^{*},\frak g):x(\infty )\in \frak b^{-}\}\oplus \Bbb C
e_{-\theta}z\tag 5.4$$
is a parabolic subalgebra, with a semidirect
decomposition
$$i_0(sl(2,\Bbb C))\propto \{x\in H^0(D^{*},\frak g):x(\infty )\in
\frak b^{-},h_{\theta}^{*}(x(\infty ))=0,Z_1=0\}.\tag 5.5$$
(this is explained in more abstract terms in Part I of
[Pi1]).  By factoring $g_{-}$ along this decomposition, we can
reduce the calculation to one inside $i_0(Sl(2,\Bbb C))$.

We therefore organize $g'$ as
$$g'=(\tilde {g}_{-}\tilde {g}_{-}(\infty )^{-1})\cdot (\tilde {g}_{
-}(\infty )g_0)\cdot (g_0^{-1}exp(-\frac b{a+bZ}e_{-\theta}z)g_0g_{
+}),\tag 5.6$$
where
$$\tilde {g}_{-}=hg_{-}exp(-\frac b{a+bZ_1}e_{-\theta}z)=\tag 5.7$$
$$\{exp(-a^{-1}\bar {b}e_{\theta}z^{-1})a^{-h_{\theta}}\}\{exp(a^{
-1}be_{-\theta}z)g_{-}exp(-Z_1e_{\theta}z^{-1})exp(-a^{-1}be_{-\theta}
z)\}$$
$$\{exp(a^{-1}be_{-\theta}z)exp(Z_1e_{\theta}z^{-1})exp(-\frac b{
a+bZ}e_{-\theta}z)\}.\tag 5.8$$
We have organized the factors (in braces) to show that
$\tilde {g}_{-}\in G(\Bbb C((z^{-1})))$.

We calculate
$$\tilde {g}_{-}(\infty )=a^{-h_{\theta}}e^{a^{-1}b[e_{-\theta},x_
1']+\frac 12(a^{-1}b)^2[e_{-\theta},[e_{-\theta},x_2]]}(1+a^{-1}b
Z_1)^{-h_{\theta}}$$
$$=e^{ab[e_{-\theta},a^{-ad(h_{\theta})}(x_1')]+b^2Z_2e_{-\theta}}
(a+bZ_1)^{-h_{\theta}}.\tag 5.9$$
When we plug this into $(5.6)$, we obtain part (a).

Part (b) is proven in the same way.  \qed
\enddemo

We now focus on the left action of $h$ on $g_{-}$, and seek
the analogue of Lemma $(3.13)$.

Suppose that $\pi =\pi_{\lambda}$ is an irreducible highest weight
representation of $G$, corresponding to the dominant
integral weight $\lambda$.  We will write $\pi_{*}$ for the Lie algebra
representation.  We can then write
$$\pi (g_{-})=1+g_1z^{-1}+g_2z^{-2}+..$$
$$=1+\pi_{*}(x_1)z^{-1}+(\pi_{*}(x_2)+\frac 12\pi_{*}(x_1)^2)z^{-
2}+..\tag 5.10$$
where the $g_n\in \Cal L(V(\pi ))$.  We will write $g_0=1$.

To generalize the discussion in \S 3 as directly as possible,
we suppose that we can choose $\lambda$ so that $\lambda (h_{\theta}
)=1$ (for
the classical algebras the $\lambda$ corresponding to the defining
representation has this property).  Let $v_{-\lambda}$ denote a
lowest weight vector.  The assumption $\lambda (h_{\theta})=1$ implies
that $v_{-\lambda}$ and $\pi_{*}(e_{\theta})v_{-\lambda}$ span a subrepresentation for
$i_0(sl(2,\Bbb C))$ ($\simeq sl(2,\Bbb C)$) isomorphic to the defining
representation.

Let $D$ and $B$, ($D_n$ and $B_n$, resp.)  denote the functions of
$g_{-}$ ($g_n$, resp.)  which pick out the $v_{-\lambda}$ and $\pi_{
*}(e_{\theta})v_{-\lambda}$
components of $\pi (g_{-})v_{-\lambda}$ ($\pi (g_n)v_{-\lambda}$, resp.),  where we split
off the other components using the essentially unique
$K$-invariant Hermitian form.  Note that $D$ and $B$ are
formal power series in $z^{-1}$, and $D_0=1$, since $g_0=1$ by
convention.

Because $v_{-\lambda}$ is lowest weight, and $l_0$ and $e_{-\theta}$ are lower
triangular, using $(a)$ of Lemma $(5.2)$, we obtain
$$\pi (g_{-}')(v_{-\lambda})=\pi (h)\pi (g_{-})(a+bZ_1)^{\pi_{*}(
h_{\theta})}(v_{-\lambda})\tag 5.11$$
$$=(a+bZ_1)^{-1}\pi (h)(\sum_{n=0}^{\infty}g_nv_{-\lambda}z^{-n})
.\tag 5.12$$
This implies the following generalization of $(3.13)$.

\proclaim{Lemma (5.13)}For the left action of $h$, as in (a)
of Lemma $(5.2)$, we have
$$\left(\matrix B(g_{-}')\\
D(g_{-}')\endmatrix \right)=(a+bZ_1)^{-1}\left(\matrix d&cz^{-1}\\
bz&a\endmatrix \right)\left(\matrix B(g_{-})\\
D(g_{-})\endmatrix \right),$$
hence for $n\ge 1$,
$$\left(\matrix B_n(g_{-}')\\
D_{n-1}(g_{-}')\endmatrix \right)=(a+bZ_1)^{-1}\left(\matrix dB_n
(g_{-})+cD_{n-1}(g_{-})\\
bB_n(g_{-})+aD_{n-1}(g_{-})\endmatrix \right),$$
and
$$B_n'(g_{-}')=\frac {c+dB'_n(g_{-})}{a+bB'_n(g_{-})},$$
where $B_1'=B_1$, $B_n'=B_n/D_{n-1}$, $n>1$.
\endproclaim

\demo{Proof of Theorem $(0.1)$} We first suppose that
$\frak g\ne E_8$.  Assuming this, we can find a representation $\pi$,
as above, such that $\lambda (h_{\theta})=1$; for the classical algebras
we can choose the defining representation, and for the
exceptional algebras, other than $E_8$, we can choose the
smallest nontrivial representation.

We can now proceed as in the $SU(2)$ case.  We again
introduce the Wiener measures parameterized by inverse
temperature $\beta$.  By asymptotic invariance and the
$i_0(SU(2))$-equivariance of $B_n'$ in Lemma $(5.13)$, each of the
$B_n'$ has the canonical invariant distribution,
$\Cal Z^{-1}(1+\vert B_n'\vert^2)^{-2}dm(B_n')$, in the limit $\beta
\to 0$, and hence each is
tight.  The tightness of $B_1$, and the irreducible action of
$K$ on $\frak g\simeq \{\theta_1\}$, implies that $\theta_1=g_1$ is tight.  Thus $
D_1$ is
tight.  Lemma $(5.13)$ now implies that $B_2$ is tight.  This
implies that the analogous coefficient of $\theta_2$ is tight.
The action of $K$ now implies that $\theta_2$ is tight, and so on.

To avoid excluding $E_8$, we consider the adjoint
representation in place of $\pi$ above.  We then consider
the three-dimensional $i_0(sl(2,\Bbb C))\subset \frak g$ in place of
$span\{v_{-\lambda},\pi_{*}(e_{\theta})v_{-\lambda}\}$, and an affine coordinate analogous to
$B_n'$ for the corresponding 2-sphere of (projective images
of) lowest root vectors for $sl(2,\Bbb C)$ inside $\Bbb P(i_0(sl(
2,\Bbb C)))$.

One approach to the tightness of $g_0$ is given in \S 4.3 of
Part III of [Pi1].  Here we will consider an argument
akin to that in \S 3.  This will apply provided
$\frak g\ne G_2,F_4,E_8$.  At the end we will give another argument
that works in general.

To prove that $g_0$ is tight, it suffices to show that the
projection of $g_0$ into $G$ modulo a finite subgroup is
tight.  Thus it suffices to show that $\pi (g_0)$ is tight, for
some irreducible (not necessarily faithful)
representation $(\pi ,V(\pi ))$ of $G$.  To show
this, we claim it suffices to prove that a single matrix
coefficient $\langle\pi (g_0)v,w\rangle$ is tight, for some $v,w\in
V(\pi )$.  To
see this, note that $K\times K$ leaves each $\pi (g_0)_{*}\nu_{\beta}$ invariant,
and acts irreducibly on $\Cal L(V(\pi ))$.  Thus if one matrix
coefficient is tight, then all matrix coefficients are
tight, because the span of the orbit of this single
coefficient will be the entire linear dual of $\Cal L(V(\pi ))$.  This
implies that $\pi (g_0)$ will be tight, viewed as a map into
$\Cal L(V(\pi ))$ [Thus in the limit $\beta\to 0$, we will definitely obtain
$K\times K$-invariant probability measures on this space of
matrices, although we do not a priori know they are
supported on $\pi (G)$].  Since $\pi (G)$ is an algebraic group, it
is cut out by polynomial equations in the matrix
coefficients, and hence these polynomials represent tight
variables.  Since these equations are satisfied for each $\beta$,
they will be satisfied in the limit.  Thus $\pi (g_0)$ will be
tight, viewed as a map into $\pi (G)$, and hence $g_0$ will be
tight.

With probability one, for each $\beta$, we can uniquely factor
$g_0$, $g_0=l_0m_0a_0u_0$, where $l_0\in N^{-}$, $u_0\in N^{+}$, $
m_0\in T$, and
$a_0\in exp(\frak h_{\Bbb R})$ (by Corollary $(1.14)$).  The factors $
l_0$ and $u_0$
are affine coordinates for the flag spaces $K/T\simeq G/B^{+}$
and $T\backslash K\simeq B^{-}\backslash G$, respectively.  Since each Wiener
measure is $K\times K$-invariant, $l_0$ and $u_0$ have the unique
$K$-invariant distributions for all $\beta$, hence are tight.
Similarly $m_0$ has the uniform distribution for all $\beta$.  The
key issue concerns the noncompact directions, $a_0$.

In the $SU(2)$ and $S^2$ cases we proved the tightness of
$a_0$ by observing that $1/a_0$ has the same limiting
distributional properties as $a_0$.  A generalization of this
is the following

\proclaim{Lemma (5.14)} Write $a_0=\prod_1^ra_j^{h_j}$, where the $
h_j$
are the coroots of the simple positive roots $\alpha_j$, and $r$
denotes the rank of $\frak g$.  Suppose that $\frak g\ne G_2,F_4,
E_8$.  Then
for some pair $i,j$, $a_i$ and $1/a_j$ will have the same
limiting distribution properties with respect to the $\nu_{\beta}$.
\endproclaim

These symmetries arise from outer automorphisms
(parameterized by $C(K)$, which is vacuous in the excluded
cases).  This is discussed in Appendix B.

Now suppose that $g_0$ is not tight.  Then for each $j$, all
the matrix coefficients for $\pi_{\Lambda_j}$ will not be tight, and in
particular each $a_j$ will not be tight.  The Lemma implies
that for some $j$, $1/a_j$ is not tight, i.e.  there exists
$\epsilon >0$ such that for all $k$, there is $\beta_k$ ($\to 0$) such that
$\nu_{\beta_k}\{a_j<\frac 1k\}>\epsilon$.

We claim that this implies that positive probability
accumulates on the lower strata of $G$, as $\beta\to 0$.  The
argument is essentially the same as in \S 3.  We can
reduce to that situation using the following procedure
(which will be used again in the proof of Lemma $(6.5)$
below).  Introduce the parabolic subgroup $P_{(j)}$ of $G$
which corresponds to the root $\alpha_j$.  Let $P_{(j)}^{+}$ denote the
nilradical, and let $P_{(j)}^{-}$ denote the opposite nilpotent
subalgebra.  We then consider the
$i_{\alpha_j}(SL(2,\Bbb C))$-equivariant projection $G\to P_{(j)}^{
-}\backslash G/P_{(j)}^{+}$.  This
projection is injective on $i_{\alpha_j}(SL(2,\Bbb C))$ itself.  When we
project the $\nu_{\beta}$ to this space, we obtain
$SU(2)\times SU(2)$-invariant measures on $SL(2,\Bbb C)$ (identified
with the image of $i_{\alpha_j}$).

This now leads to a contradiction with the following
Lemma, which will complete the proof of Theorem $(0.1)$,
at least in the cases $\frak g\ne G_2,F_4,E_8$.

\proclaim{Lemma (5.15)} Suppose that $\{\lambda_{\beta}:\beta >0\}$ is a
family of probability measures on $SL(2,\Bbb C)$
quasi-invariant and asymptotically invariant with
respect to $SU(2)\times SU(2)$.  Then $1/a$, as a map into $\Bbb R$, is
tight.
\endproclaim

\demo{Proof of (5.15)} By way of contradiction, suppose
that $1/a$ is not tight.  Then there exists $\epsilon >0$ such that
for all $k$, there is $\beta_k$ ($\to 0$) such that $\lambda_{\beta_
k}\{a_j<\frac 1k\}>\epsilon$.

When we factor
$$\left(\matrix a&b\\
c&d\endmatrix \right)=\left(\matrix 1&0\\
z&1\endmatrix \right)\left(\matrix a&0\\
0&a^{-1}\endmatrix \right)\left(\matrix 1&w\\
0&1\endmatrix \right),\tag 5.16$$
for each $\beta$, $z$ and $w$ will have the standard
$\Cal Z^{-1}(1+\vert\zeta\vert^2)^{-2}dm(\zeta )$ distributions, in the limit $
\beta\to 0$.  Thus
there is an $M=M_{\epsilon}$ such that $\vert z\vert ,\vert w\vert
<M$ with
$\lambda_{\beta}$-probability $>1-\epsilon /3$ for all sufficiently small $
\beta$.  Since
$b=aw$, $c=az$, these will accumulate positive probability
as $\beta\to 0$.

To control $d$ we need to write down an analogue of $\zeta$
from \S 4.  This is straightforward, but somewhat messy.
We write
$$\left(\matrix a&b\\
c&d\endmatrix \right)=k_1p,\quad k_1\in S(2,\Bbb C),\quad p>0,\tag 5.17$$
$$p^2=\left(\matrix A&B\\
\bar {B}&D\endmatrix \right)=\left(\matrix \vert a\vert^2+\vert c
\vert^2&\bar {a}b+\bar {c}d\\
a\bar {b}+c\bar {d}&\vert b\vert^2+\vert d\vert^2\endmatrix \right
).\tag 5.18$$
There is a $k_2\in SU(2,\Bbb C)$ which will diagonalize $p$, or
equivalently $p^2$.  We seek an affine coordinate $\zeta$ for
$k_2U(1)\in SU(2)/U(1)$.  This will represent an equivariant
map to the $2$-sphere for the diagonal copy of $SU(2)$.

The eigenvalues of $p^2$ are $\lambda =(A+D\pm ((A+D)^2-4))^{1/2}
/2$
Then $p^2-\lambda$ equals
$$\left(\matrix \frac {A-D}2\pm\sqrt {(\frac {A+D}2)^2-1}&B\\
\bar {B}&\frac {D-A}2\pm\sqrt {(\frac {A+D}2)^2-1}\endmatrix \right
).\tag 5.19$$
We can take
$$\zeta =\frac {\bar {B}}{D-A+\sqrt {(A+D)^2-4}}=\frac {a\bar {b}
+z(1+bc)}{D-A+\sqrt {(A+D)^2-4}},\tag 5.20$$
where we used $ad-bc=1$ to rewrite $\bar {B}$ in terms of $z$
(from $(5.16)$) and other variables not including $d$.  We
then have
$$D-A+\sqrt {(A+D)^2-4}=(a\bar {b}+z(1+bc))/\zeta\tag 5.21$$
Since $D\ge 0$, we obtain
$$\sqrt {(A+D)^2-4}\le (a\bar {b}+z(1+bc))/\zeta ,\tag 5.22$$
implying
$$(A+D)^2\le 4+\{(a\bar {b}+z(1+bc))/\zeta \}^2\tag 5.23$$
In this expression, only $D=\vert b\vert^2+\vert d\vert^2$ involves $
d$.

Since $\zeta$ represents an equivariant map to the $2$-sphere
for the diagonal $SU(2)$, $\zeta$ has the usual invariant
distribution in the limit $\beta\to 0$.  Hence we can find
$M'=M'_{\epsilon}$ such that such that $\vert 1/\zeta\vert <M'$ with
$\lambda_{\beta}$-probability $>1-\epsilon /3$ for all sufficiently small $
\beta$.  We
now see that with positive probability $b$, $c$, and $d$ will
remain finite with positive probability as $k\to\infty$.  This
means that we do obtain a $SU(2)$-invariant measure on
the lower strata, in the limit $k\to\infty$, and this contradicts
$(1.14)$.  \qed
\enddemo

If $\frak g=G_2,F_4$ or $E_8$, then this argument does not work,
because we cannot apply $(5.14)$.  In general we consider
$a_0^{-\theta}$.  This is the ``$\vert a\vert^2$'' variable for $
i_0(SL(2,\Bbb C))$; this is
due to the minus sign appearing in front of $h_{\theta}$ in $(5.1
)$
(see $(6.11)$ below).  If none of the $a_j$ are tight, then $a_0^{
-\theta}$
tends to zero with some positive probability uniform in
$\beta$.  The Lemma $(5.15)$ then applies and we obtain a
contradiction.  \qed
\enddemo

\bigskip

\centerline{\S 6. General Symmetric Space Case.}

\bigskip

We continue to assume that $\frak g$ is simple (so that we will
be considering the type I symmetric space case).  The
homomorphism $i_0$ in $(5.1)$ will continue to play a critical
role.  Because $\Theta$ preserves the triangular decomposition
of $\frak g$, $\Theta (\theta )=\theta$, and $\Theta (\frak g_{\theta}
)=\frak g_{\theta}$, hence $\Theta (e_{\theta})=-\epsilon e_{\theta}$, where
$\epsilon =\pm 1$ (it appears that both possibilities are unavoidable;
for example for $Sp(n)/U(n)$, it appears that we must
have $\epsilon =+1$, and for the real Grassmannian of oriented
$p$-planes in $\Bbb R^{p+q}$, it appears we must have $\epsilon =
-1$).
Thus the extension of $\Theta$ to loop space will map the
image of $i_0$ into itself, and when $\epsilon =1$, this action will
be isomorphic to $(4.2)$ via $i_0$, otherwise it is trivial.

\proclaim{Lemma (6.1)} Suppose that $\left(\matrix a&b\\
c&d\endmatrix \right)\in SL(2,\Bbb C)$, and
let $h=i_0(\left(\matrix a&b\\
c&d\endmatrix \right))$.  Suppose also that $g=g_{-}\cdot g_0\cdot
g_{+}$ is a
point in the top stratum of $\bold L(G/G_{\Bbb R})$, so that
$$g_0^{*\Theta}=g_0,\quad and\quad g_{+}=g_{-}^{*\Theta}.$$
We write $g_{-}$ as in $(5.3)$, and $x_n=x_n'+Z_ne_{\theta}$.  Let
$$Z_1'=Z_1-\frac {\epsilon\bar {b}}{\bar {a}+\epsilon\bar {b}\bar {
Z}_1}\langle Ad(g_0)e_{\theta},e_{\theta}\rangle /\langle e_{\theta}
,e_{\theta}\rangle .$$
If both $a+\epsilon bZ_1$ and $a+bZ_1'$ are not zero, then the point
$$g'=h\cdot \{g_{-}\cdot g_0\cdot g_{+}\}\cdot h^{*\Theta}$$
is also in the top stratum, and the triangular
factorization $g'=g_{-}'\cdot g_0'\cdot g_{+}'$ is given by
$$g_{-}'=hg_{-}g_0exp(\frac {-\epsilon\bar {b}}{\bar {a}+\epsilon
\bar {b}\bar {Z}_1}e_{\theta}z^{-1})g_0^{-1}exp(\frac {-b}{a+bZ_1'}
e_{-\theta}z)(a+bZ_1')^{h_{\theta}}l_0^{\prime -1},$$
$$g_0'=l_0'(a+bZ_1')^{-h_{\theta}}g_0(\bar {a}+\epsilon\bar {b}\bar {
Z}_1)^{-h_{\theta}}u_0,$$
$$u_0=exp(-ac[e_{\theta},a^{-ad(h_{\theta})}(y_1')]-c^2W_2e_{\theta}
)\in N^{+},$$
and $l_0'$ is lower triangular.
\endproclaim

\flushpar Remark (6.2).  Note that for generic $g_0$, with
triangular decomposition $g_0=l_0diag(g_0)u_0$, we can write
$$\langle Ad(g_0)e_{\theta},e_{\theta}\rangle /\langle e_{\theta}
,e_{\theta}\rangle =diag(g_0)^{\theta},\tag 6.3$$
and
$$a+bZ_1'=\frac {\vert a+bZ_1\vert^2+\epsilon b\bar {b}diag(g_0)^{
\theta}}{\bar {a}+\epsilon\bar {b}\bar {Z}_1},\tag 6.4$$
These formulas specialize to those in Lemma $(4.4)$ (with
$\epsilon =1$).

\demo{Proof of (6.1)}This formula is obtained from Lemma
$(5.2)$ by first applying $h^{*\Theta}$ on the right, then $h$ on the
left (one obtains formulas with considerably different
appearance if one does this in the opposite order, and
then it is not so easy to compare with $(4.4)$).  \qed\
\enddemo

\demo{Proof of Theorem (0.13)} Let $\nu_{\beta}$ denote Wiener
measure for $L_{C^0}U$ with inverse temperature $\beta$, and let
$\tilde{\nu}_{\beta}$ denote the projection to $L_{C^0}(U/K)$; we view $
\tilde{\nu}_{\beta}$ as a
probability measure on $\bold L(G/G_{\Bbb R})$.  Because the map
$LU\to L(U/K)$$:g\to gg^{-\Theta}$ is equivariant for the left actions
of $L_{pol}U$, the measures $\tilde{\nu}_{\beta}$ form an asymptotically
invariant family for this action.

\proclaim{Lemma (6.5)}The top stratum
$\bold L(G/G_{\Bbb R})\cap\tilde{\Sigma}_1^{formal}$ has full measure with respect to $
\tilde{\nu}_{\beta}$.
\endproclaim

\demo{Proof of $(6.5)$} The argument is the same as in
Proposition $(2.1.1)$ of Part I of [Pi1], for which it is
convenient to use Kac-Moody technology.  In the proof,
for notational clarity, we will suppose $\Theta (e_{\theta})=-e_{
\theta}$; we
will note the minor changes necessary in the opposite
case at the end of the proof.

There is a transitive action
$$G(\Bbb C((z^{-1})))\times \bold L(G/G_{\Bbb R})\to \bold L(G/G_{
\Bbb R}):g,h\to g\cdot h\cdot g^{*\Theta}.\tag 6.6$$
This action is well-defined, because if $g\in G(\Bbb C((z^{-1})))$,
then $g^{*\Theta}\in G(\Bbb C((z)))$, and these groups act from the left
and right of $\bold LG$, respectively.  We let $\Cal B^{\pm}$ denote the
upper and lower Borel subgroups of $G(\Bbb C((z^{-1})))$,
respectively.

The decomposition for $\bold LG$ strictly analogous to $(1.7)$ is
given by
$$\bold LG=\bigsqcup_{W\alpha Hom(S^1,T)}\tilde{\Sigma}_w^L,\quad
where\quad\tilde{\Sigma}_w^L=\Cal B^{-}w\Cal B^{+},\tag 6.7$$
and the superscript $L$ is there to remind us of loop
space.  This induces a decomposition
$$\bold L(G/G_{\Bbb R})=\bigsqcup S_w,\quad where\quad S_w=\tilde{
\Sigma}_w^L\cap \bold L(G/G_{\Bbb R}).\tag 6.8$$
We are mainly interested in $S_1$, the ``top stratum'',
which consists of $g$ in the formal completion that can be
written uniquely as $g=l\cdot h\cdot l^{*\Theta}$, where
$l\in \Cal N^{-}=G(\Bbb C[[z^{-1}]])_1$ (the formal completion of the
profinite nilpotent algebra spanned by the negative roots
of $L_{pol}\frak g$), and $h=h^{*\Theta}\in H$.  To say that $g$ belongs to $
S_1$
is thus equivalent to the two conditions $g\in\tilde{\Sigma}^{for
mal}_1$ and
$g_0\in\tilde{\Sigma}_1$.

Let $\Cal P^{-}_{(0)}$ denote the parabolic subgroup, with Levi
decomposition,
$$\Cal P^{-}_{(0)}=i_0(SL(2,\Bbb C))\Cal B^{-}=i_0(SL(2,\Bbb C))\alpha
\Cal R^{-}_{(0)}.\tag 6.9$$
$(\Cal R^{-}_{(0)}$ is the subgroup of $\Cal N^{-}$ that corresponds to the span
of the negative root spaces, $\Bbb Ce_{\theta}z^{-1}$ excluded).

We consider the diagram
$$\matrix \bold L(G/G_{\Bbb R})&@>{q}>>&\Cal R_{(0)}^{-}\backslash
\bold L(G/G_{\Bbb R})\\
\uparrow&&\uparrow\\
S_1\cup w_{(0)}S_1&\to&i_0(\phi (SL(2,\Bbb C)/SU(1,1)))\\
\uparrow&&\uparrow\\
S_1&\to&i_0(\tilde{\Sigma}_1)\endmatrix \tag 6.10$$
where $w_{(0)}=i_0(\left(\matrix 0&-1\\
1&0\endmatrix \right))$.  The map $q$ is $i_0(SL(2,\Bbb C))$
equivariant, where the actions in question are induced
by $(6.6)$, because this group normalizes $\Cal R^{-}_{(0)}$.  The up
arrows are all injective (the only question here might
involve the upper right up arrow; if $g,h\in i_0(SL(2,\Bbb C))$,
$gg^{\tau}=1$, $hh^{\tau}=1$, and $g=lhl^{*\Theta}$, for some $l\in
\Cal R^{-}_{(0)}$, then
$gl^{\tau}=lh$, an equality in $\Cal P^{-}_{(0)}$; when we quotient out by
$\Cal R^{-}_{(0)}$, we obtain $g=h$).

The sets on the left of $(6.10)$ are the full $q$-inverse
images of the sets on the right.  For if $g$ is in the
formal completion and can be written uniquely as
$l\cdot h\cdot l^{*\Theta}$, we can factor $l=rl_0$ along the Levi
decomposition, and then $g'=l_0hl_0^{*\Theta}\in i_0(SL(2,\Bbb C)
)$ and
$g'g^{\prime\tau}=1$.  Conversely if we have a factorization for $
g'$,
we clearly obtain an element in $S_1$.

Now suppose that $\nu$ is a $L_{pol}K$-quasi-invariant
probability measure on the formal completion.  Consider
$q_{*}\nu$ restricted to $i_0(\phi (SL(2,\Bbb C)/SU(1,1))$.  This measure
(which may have total mass less than one) is
$i_0(SU(2))$-quasi-invariant.  By Corollary $(1.14)$, it is
supported on the top stratum.  Therefore the top
stratum and its $w_{(0)}$-translate are equivalent, up to a
set of $q_{*}\nu$-measure zero.  Together with the fact that
the sets on the left of $(6.10)$ are the full $q$-inverse
images of the sets on the right, this implies that the
$w_{(0)}$-translate of $S_1$ and $S_1$ are equivalent up to a set of
$\nu$-measure zero.

Now recall that $S_1$ consists of $g\in\tilde{\Sigma}_1^{formal}\cap
\bold L(G/G_{\Bbb R})$
such that $g_0$ is in the top stratum of $G/G_{\Bbb R}$.  By
Corollary $(1.14)$, applied to the $U$-quasi-invariant measure
$g_{0*}\nu$, we conclude that $S_1$ and the top stratum are the
same, up to a set of $\nu$-measure zero.

The top stratum is fixed by $W$.  The group generated by
$w_{(0)}$ and $W$ is the affine Weyl group, $W\alpha Hom(S^1,T)$).
Therefore the translates of the top stratum for the
formal completion by the affine Weyl group are all
equivalent, up to $\nu$-measure zero.  But the union of
these translates is the whole space.  Therefore the
complement of the top stratum must have $\nu$-measure
zero.

In the event that $\Theta (e_{\theta})=e_{\theta}$, we replace $S
L(2,\Bbb C)/SU(1,1)$
by $SL(2,\Bbb C)/SU(2)$, and the same argument applies.  \qed
\enddemo

We now turn to the proof that $g_0$ is tight.  With $\nu_{\beta}$
probability one, by $(1.14)$, we can write $g_0=l_0\bold wma_0l_0^{
*\Theta}$,
as in $(d)$ of $(1.8)$.  As in the group case it suffices to
show that in some representation, there is some matrix
coefficient which is tight.  If we write $a_0=\prod a_j^{h_j}$, it
suffices to show one of the $a_j$ is tight (as a function
into $\Bbb R$).

As something of a digression, consider first a case in
which $\Theta$ is an inner automorphism (this is equivalent to
$\frak a_0=0$).  In this case the action of $\Theta$ on $C(K)$ and $
\frak h$ is
trivial.  In the notation of Appendix B, the corresponding
automorphism, $\sigma_{\Delta}$, will map $\bold L(G/G_{\Bbb R})$ into itself.  Lemma
$(5.14)$ applies.  Thus if all of the $a_j$ are not tight, then
for some $i$, $1/a_i$ is not tight.  This leads to a
contradiction, using either $(4.22)$ or $(5.15)$, depending upon
whether $\Theta (e_j)=\pm e_j$.

Now suppose that $\Theta$ is an outer automorphism.  In some
cases (e.g.  $SU(2n)/SO(2n)$), there still exist nontrivial
$\Delta\in C(K)$ fixed by $\Theta$. This is illustrated in Appendix B.

In general we argue as follows.  If all of the $a_j$ are not
tight, this implies that $a_0^{\theta}$ is not tight.  Thus $a_0^{
-\theta}$ tends
to zero with some positive probability uniform in $\beta$.
But this is the ``$\vert a\vert$'' variable for $i_0(SL(2,\Bbb C)
)$:  given an
element in $SL(2,\Bbb C)$ as in $(5.16)$, when we apply $i_0$, as
defined by $(5.1)$,
$$i_0(\left(\matrix a&0\\
0&a^{-1}\endmatrix \right))=a^{-h_{\theta}}\quad\implies\quad\vert
a\vert^2=a_0^{-\theta}.\tag 6.11$$
We now consider the diagram as in $(6.10)$.  When we
push our measures forward, we obtain an asymptotically
invariant family of quasi-invariant measures for $SU(2)$
acting on $SL(2,\Bbb C)/SU(1,1)$ or $SL(2,\Bbb C)/SU(2)$ (using the
isomorphism $i_0$).  We then apply Lemmas $(4.22)$ and
$(5.15)$, respectively.  This leads to a contradiction.

Thus we know that $g_0$ is tight.  To complete the proof
we need to show that the coefficients of $\theta_{-}$, or
equivalently the coefficients of $x$ (where $g_{-}=exp(x)$),
are tight.  The argument is the same as the first
argument we gave in the $S^2$ case.

We consider the action of $h$ in Lemma $(6.1)$.  The
formula for $g_0'$ in Lemma $(6.1)$ implies
$$\langle Ad(g_0')e_{\theta},e_{\theta}\rangle =(a+bZ_1')^{-2}(\bar {
a}+\epsilon\bar {b}Z_1)^{-2}\langle Ad(g_0)e_{\theta},e_{\theta}\rangle\tag 6.12$$
$$=\frac 1{(\vert a+bZ_1\vert^2+\epsilon\vert b\vert^2\langle Ad(
g_0)e_{\theta},e_{\theta}\rangle )^2}\langle Ad(g_0)e_{\theta},e_{
\theta}\rangle ,\tag 6.13$$
where the second equality uses $(6.4)$.  We have already
proven that $g_0$ is tight, so that $\langle Ad(g_0)e_{\theta},e_{
\theta}\rangle$ is tight.
Asymptotic invariance implies that $g_0'$ (for a given $h$) is
also tight.  If we take $h$ with $a=0$ and $b=1$, we then
see that $Z_1$ is tight.

Now $Z_1$ is the $e_{\theta}$ component of $x_1$.  The constants $
K$ act
irreducibly by the adjoint representation on $\{x_1\}$.  Since
the constants leave the measures $\nu_{\beta}$ invariant, it follows
that $x_1$ is tight.

We then apply the formula for $g_{-}'$ to compute the
adjoint matrix coefficient (where $Ad$ is implicit, and we
set $\beta =-\epsilon\bar {b}(\bar {a}+\bar {b}\bar {Z}_1)^{-1}$)
$$\langle Ad(g_{-}')e_{-\theta},e_{-\theta}\rangle =\langle hg_{-}
g_0exp(\beta e_{-\theta}z^{-1})g_0^{-1}(a+bZ'_1)^{-2}e_{-\theta},
e_{-\theta}\rangle )\tag 6.14$$
We again take $h$ with $a=0$, $b=1$.  We move $h$ to the
right side of the inner product.  In the adjoint
representation for $i_0$ we calculate that $h^{-1}e_{-\theta}=-z^{
-2}e_{\theta}$.
Also $ad(e_{-\theta})^3=0$.  We then see that $(6.14)$ equals
$$(Z_1')^{-2}\langle g_{-}exp(\beta e^{g_0}_{-\theta}z^{-1})e_{-\theta}
,-z^{-2}e_{\theta}\rangle )\tag 6.15$$
$$=(Z'_1)^{-2}\langle (\sum_{n=0}^{\infty}g_nz^{-n})(z^2+\beta ad
(e_{-\theta}^{g_0})z+\frac 12\beta^2ad(e_{-\theta}^{g_0})^2)e_{-\theta}
,e_{\theta}\rangle .\tag 6.16$$
where we are doing this calculation for $z\in S^1$, so that
$z^{-1}=\bar {z}$ (these calculations closely mirror, but do not
quite reduce to, the calculations $(4.5)-(4.13)$, because we
are considering the adjoint representation, rather than
the analogue of the defining representation).  We now
expand the left hand side of $(6.11)$ and $(6.16)$ in powers of
$z^{-1}$.  By isolating the terms of order $-n$, we obtain
$$\langle g_n'e_{-\theta},e_{-\theta}\rangle =(\langle g_{n+1}e_{
-\theta},e_{\theta}\rangle +..)/(Z'_1)^2,\tag 6.17$$
where the trailing terms involve $g_m$, $m\le n$.

We know that $Z_1'$ and $g_1=x_1$ are tight.  This implies
that one matrix coefficient for $g_2$ is tight, and hence
that a matrix coefficient for $x_2$ is tight.  Using the
action of the constants $K$, we conclude $x_2$ is tight.  We
then continue in this same way.  \qed
\enddemo

\bigskip

\centerline{\S 7. Diagonal Distribution Conjecture.}

\bigskip

For $g\in\phi (G/G_{\Bbb R})\cap\tilde{\Sigma}_1$, we can write $
g=l\bold wma_{\phi}l^{*\Theta}$, as in
(d) of $(1.8)$.  Let $\tilde{\Sigma}_{\bold w}$ denote the set of $
g$ in this
intersection for fixed $\bold w\in T_0^{(2)}/T_0^{(2)}\cap exp(i\frak a_
0)$.  The
following is proven in [Pi4].

\proclaim{Proposition (7.1)} Suppose that $U/K$ is
Hermitian symmetric or a group.  Then for $\lambda\in (i\frak t_0
)^{*}$,
$$\int_{\phi (U/K)\cap\tilde{\Sigma}_{\bold w}}a_{\phi}(g)^{-i\lambda}
=\frac 1M\prod^{\bold w}\frac {\langle\delta ,\alpha\rangle}{\langle
\delta -i\lambda ,\alpha\rangle}$$
where $\delta$ is half the sum of the positive complex roots,
and the product is over positive roots which are
imaginary and noncompact with respect $Ad(\bold w)\Theta$.
Therefore
$$\int_{\phi (U/K)}a_{\phi}(g)^{-i\lambda}=\frac 1{\#\{[\bold w]\}}
\sum_{[\bold w]}\prod^{\bold w}\frac {\langle\rho ,\alpha\rangle}{
\langle\delta -i\lambda ,\alpha\rangle}$$
where the sum is over equivalence classes in
$T_0^{(2)}/T_0^{(2)}\cap exp(i\frak a_0)$ such that $\bold w\in\phi
(U/K)$.
\endproclaim

\flushpar Example $(7.2)$.  We write $g\in\phi (S^2)$ as
$g=\left(\matrix a_0&b_0\\
-\bar {b}_0&a_0\endmatrix \right)$, where $a_0\in \Bbb R$, $a_0^2
+\vert b_0\vert^2=1$.  We have
$a_{\phi}=\left(\matrix \vert a_0\vert&0\\
0&\vert a_0\vert^{-1}\endmatrix \right)$, and $\bold w$ is $\pm 1$, corresponding to the
upper and lower hemispheres.  Write $\lambda =\lambda\alpha_1$, where
$\alpha_1=\lambda_1-\lambda_2$ is unique noncompact root.  We calculate
$$\int_{\phi (S^2)\cap\tilde{\Sigma}_{\pm}}a_{\phi}^{-i\lambda}=\int_
0^1t^{-i2\lambda}dt=\frac 1{1-i2\lambda}=\frac {\langle\frac 12\alpha_
1,\alpha_1\rangle}{\langle\frac 12\alpha_1-i\lambda\alpha_1,\alpha_
1\rangle}\tag 7.3$$
(and note $\rho =\alpha_1$).  The fact that the integrals
corresponding to $\bold w=\pm 1$ are the same is peculiar to this
example.

\smallskip

Now suppose that we consider $L(U/K)$.  Let $\nu$ denote an
$L_{pol}U$-invariant probability measure on the formal
completion.  We think of this loop space as a symmetric
space, with $\frak k$ replaced by $L\frak k$, $\frak p$ by $L\frak p$, and so on.  We
then write down the analogue of the formula in $(7.1)$ (for
the group case, see $\S 4.3$ of Part III of [Pi1]).  Using the
standard product formula for the $sine$ function, we
obtain the following

\proclaim{Conjecture (7.4)}Given $g$ in the top stratum of
$\bold L(G/G_{\Bbb R})$, we write $g_0=l_0\bold wma_{\phi}l_0^{*\Theta}$ as in $
(d)$ of $(1.8)$.
Then
$$\int_{\bold L(G/G_{\Bbb R})}a_{\phi}^{-i\lambda}d\nu =\frac 1{\#
\{\bold w\}}\sum_{\bold w}\prod^{\bold w}\frac {sin(\frac {\pi}{\dot {
g}}\langle\delta ,\alpha\rangle )}{sin(\frac {\pi}{\dot {g}}\langle
\delta -i\lambda ,\alpha\rangle )}$$
where here the inner product is normalized so that a
long root has length $\sqrt 2$.
\endproclaim

In the $S^2$ case it is a routine exercise to explicitly
compute $(g_0)_{*}\nu$ itself.

\proclaim{Lemma (7.5)}In the $S^2$ case, in terms of the
coordinates $(4.15)$, with $k=\left(\matrix a&b\\
-\bar {b}&\bar {a}\endmatrix \right)$, the Conjecture $(7.4)$
is equivalent to
$$(g_0)_{*}\nu =\Cal Z^{-1}sech(2x)dk\times dx$$
\endproclaim

\demo{Sketch of the calculation} By $(4.19)-(4.20)$,
$a_0=(2\vert a\vert^2-1)ch(2x)+sh(2x)$, and by $(C.6)$ of Appendix C,
the $SL(2,\Bbb C)$-invariant measure on $SL(2,\Bbb C)/SU(1,1)$ is
given by $d\eta =$$cosh^2(2x)dk\times dx$.  If $\delta (2x)d\eta$ is the $
g_0$
distribution, then we must compute
$$\int\vert a_0\vert^{-i\lambda}=\int\vert (2\vert a\vert^2-1)ch(
2x)+sh(2x)\vert^{-i\lambda}\delta (2x)ch^2(2x)dkdx$$
$$=\int_{x=-\infty}^{+\infty}\int_{u=-1}^1\vert uch(2x)+sh(2x)\vert^{
-i\lambda}ch^q(2x)dxdu\tag 7.6$$
where $u=2\vert a\vert^2-1$, and we used the fact that the
$a$-distribution of Haar measure for $SU(2)$ is Lebesgue
measure on the unit disk.  We do the $u$-integral, and we
eventually obtain
$$\frac 1{1-i\lambda}\int_{-\infty}^{\infty}e^{-i2x\lambda}e^{2x}
\delta (2x)ch^2(2x)sech(2x)dx.\tag 7.7$$
According to Conjecture $(7.5)$, this equals $1/sin(\frac {\pi}2(
1-i\lambda ))$.
When we apply the inverse Fourier transform, we find
$\delta (2x)=sech^3(2x)$.
\enddemo

In general, to obtain a formula for the $g_0$ distribution, it
is necessary to use DeLorme's inversion formula for the
``diagonal transform'' of harmonic analysis for
$U$-invariant functions on $G/G_{\Bbb R}$ (see [D] and references).
This will be considered elsewhere (in the group case one
can instead use Harish-Chandra's spherical transform,
which is more elementary).

In [Pi2], I described a possible application of the
invariant measure for $LK$ to the construction of the
state space for the two dimensional sigma model with
target $K$.  That
discussion remains essentially unchanged when $K$ is
replaced by $U/K$.  The conjectural formula above for the
$g_0$ (or zero-mode) distribution leads to the conjecture
that (in the large radius limit) the radial part of
the zero-mode Hamiltonian for
the $S^2$ sigma model is equivalent to
$$-(\frac d{dx})^2+\frac 14-\frac 34sech^2(x),\quad x\in \Bbb R\tag 3.10$$
(the zero-mode Hamiltonian is itself a $U$-invariant
Laplace type operator on $G/G_{\Bbb R}$).

One theme of this paper is that, in considering natural
measures associated to the loop space $L(U/K)$,
Riemann-Hilbert factorization is important, and when
$U/K$ is positively curved, the zero-mode becomes the
noncompact space $G/G_{\Bbb R}$.  An amusing example is
$$\matrix &{{}\atop {}}&Sp(2,\Bbb C)\\
&\nearrow&&\nwarrow\\
Sp(2,\Bbb R)&&&&Sp(2)\\
&\nwarrow&&\nearrow\\
&&K=U(2)\endmatrix \tag 1.19$$
where the compact six dimensional space $Sp(2)/U(2)$
(isomorphic to the Grassmannian of oriented $2$-planes in
$\Bbb R^5$, via an exceptional low rank isomorphism) is
embedded in the $10$ dimensional space $Sp(2,\Bbb C)/Sp(2,\Bbb R)$,
with $4$ noncompact dimensions.

In connection with sigma models, it is natural ask about
loops into other ideal spaces, such as $T$, a flat torus,
$\pi\backslash G_{\Bbb R}/K$, a compact locally symmetric space with
negative curvature, or a Ricci flat space.  In the
nonpositively curved cases, in the framework of this
paper, compactness fails.  This implies that the
existence, or lack thereof, of compactness for Wiener
measures (for ideal spaces) corresponds to asymptotic
freedom, or the lack thereof, in renormalization group
analysis of the corresponding sigma models.  Whether
there is a tangible connection is unclear, as is the
question of how to extend this to more general spaces
(or to the supersymmetric setting).

\bigskip

\centerline{Appendix A. Asymptotic Invariance of Wiener Measure}

\bigskip

In this paper we will adopt the following

\proclaim{Definition $(A.1)$}
Given a group $A$ of automorphisms of a measure space
 $(M,\Cal M)$, and a
family $\{\lambda_{\beta}:\beta >0\}$ of quasi-invariant probability
measures on $M$, this family is
asymptotically invariant if
$$\int\vert 1-\frac {d\sigma_{*}\lambda_{\beta}}{d\lambda_{\beta}}
\vert d\lambda_{\beta}\to 0\quad as\quad\beta\downarrow 0,$$
for each $\sigma\in A$.
\endproclaim

If the measures $\lambda_{\beta}$ have a weak limit (relative
to bounded continuous functions, for example, in a
topological context), then the
limit will represent an invariant object. Given
 an $A$-equivariant map $\phi :M\to N$, the family $\{\phi_{*}\lambda_{
\beta}\}$
will be asymptotically invariant.

In this appendix we will discuss a slight extension
of the asymptotic invariance results in Part III of [Pi1].

Let $X=K$, a simply connected compact Lie group, which
we view as a Riemannian symmetric space (using our
normalized $Ad(K)$-invariant inner product).  Let $\nu_{\beta}$ denote
the Wiener probability measure on $L_{C^0}X$ with inverse
temperature $\beta$.

\proclaim{Proposition (A.2)}Suppose that $\sigma\in \Cal L_{W^1}A
ut(X)_0$.
Then
$$\int\vert 1-\frac {d\nu_{\beta}(g^{\sigma})}{d\nu_{\beta}(g)}\vert
d\nu_{\beta}(g)\le 2^{3/2}\frac {p^2_{T/2}(1)}{p_T(1)}\beta^{1/2}
\Cal E(\sigma )^{1/2},$$
where $\Cal E$ denotes the kinetic energy of $\sigma$, $\frac 12\int_
0^1\vert\sigma^{-1}d\sigma\vert^2dt$,
$T=1/\beta$, and $p$ is the heat kernel.
\endproclaim\

\demo{Proof of (A.2)} This result is already known if $\sigma$
is either a left or right multiplication.  The essential
point is to treat left and right multiplication
simultaneously.  We first consider unconditioned paths.

Let $\nu =\nu^{1,*}_{\beta}$ denote the Wiener probability measure on
$Path^{1,*}_{C^0}K$.  Given $l,r\in Path^{1,*}_{W^1}K$, we compute
$$\int\vert 1-\frac {d\nu (lgr)}{d\nu (g)}\vert d\nu (g)\le\int\vert
1-\frac {d\nu (gr)}{d\nu (g)}\vert d\nu (g)+\int\vert\frac {d\nu
(gr)}{d\nu (g)}-\frac {d\nu (lgr)}{d\nu (g)}\vert d\nu (g)\tag A.3$$
$$=\int\vert 1-\frac {d\nu (gr)}{d\nu (g)}\vert d\nu (g)+\int\vert
1-\frac {d\nu (lgr)}{d\nu (gr)}\vert\frac {d\nu (gr)}{d\nu (g)}d\nu
(g)\tag A.4$$
$$=\int\vert 1-\frac {d\nu (gr)}{d\nu (g)}\vert d\nu (g)+\int\vert
1-\frac {d\nu (lg)}{d\nu (g)}\vert d\nu (g)\tag A.5$$
$$\le 2\beta^{1/2}(\Cal E(r)^{1/2}+\Cal E(l)^{1/2})\le 2^{3/2}\beta^{
1/2}\Cal E(l,r)^{1/2}\tag A.6$$
where the inequality in the fourth line uses $(4.1.11)$ of
[Pi1].

Suppose that $\Delta\in C(K)$, and $l$ and $r$ satisfy $l(t+1)=l(
t)\Delta$,
$r(t+1)=r(t)\Delta$, for all $t$.  We let $\sigma (g)=lgr^{-1}$, for
$g\in Path^{1,*}K$.  Given $k\in K$, let $\nu^{1,k}_{\beta}$ denote the Wiener
probability measure on $Path^{1,k}_{C^0}K$.  Note that $\sigma$ acts on
this space.  We have the disintegration formula
$$\nu^{1,*}_{\beta}=\int_K\nu^{1,k}_{\beta}p_T(k)dk.\tag A.7$$
The inequality $(A.3)-(A.6)$ implies
$$\int_K\{\int\vert 1-\frac {d\nu^{1,k}_{\beta}(g^{\sigma})}{d\nu^{
1,k}_{\beta}(g)}\vert d\nu_{\beta}^{1,k}\}p_T(k)dk\le 2^{3/2}(\beta
\Cal E(\sigma ))^{1/2},\tag A.8$$
where a priori we only know that $\sigma_{*}\nu_{\beta}^{1,k}$ has the same
measure class as $\nu^{1,k}_{\beta}$ for $a.e.$ $k$.  We need to translate
this statement about an average over $K$ to a statement
about each $k$, especially $k=1$.

The basic idea is that we can express $\nu^{1,1}_{\beta}$ as an integral,
$$\nu^{1,1}_{\beta}=\int_K(\nu^{1,k}_{2\beta}*\nu_{2\beta}^{1,k^{
-1}})\frac {p^2_{T/2}(k)}{p_T(k)}dk.\tag A.9$$
The argument following $(4.1.11)$ of Part III of [Pi1] applies
verbatim with $\sigma (g)$ in place of the left multiplication
$g_Lg$.  \qed
\enddemo

\flushpar Remark $(A.10)$. It is important to extend
 the asymptotic invariance result above in a number
of directions. Unfortunately the method in [Pi] depends
on the fact that the target is a group
(the methods in [MM] are more robust).
This explains why Wiener measure for the target $U/K$
has not made any appearance in this paper.

\bigskip

\centerline{Appendix B. Symmetries of the Diagonal Distribution.}

\bigskip

\flushpar {\bf The Group Case.  }

\smallskip

In this appendix it is convenient to use the Kac-Moody
extension $\Bbb C\to\hat {L}_{pol}\frak g\to L_{pol}\frak g$.  Our notation will be consistent
with [KW], which compiles numerical information we
will need.  We very briefly recall that $\frak g$, which we
assume is simple, is generated by $r=rank(\frak g)$ copies of
$sl(2,\Bbb C)$, with standard bases $h_j$, $e_j$, $f_j$, $1\le j\le
r$,
satisfying the Chevalley-Serre relations; $\hat {L}_{pol}\frak g$
is generated by $\frak g$ (or the $r$ copies of $sl(2,\Bbb C)$)
 and one additional copy of $sl(2,\Bbb C)$, with
standard basis $h_0=c-h_{\theta}$, $e_0=e_{-\theta}\otimes z$, $f_
0=e_{\theta}\otimes z^{-1}$. The
Dynkin diagram of $\hat {L}_{pol}\frak g$, which is the extended diagram for
$\frak g$, encodes the (generalized Chevalley-Serre)
relations among the different copies of
$sl(2,\Bbb C)$.

The
automorphisms of the Dynkin diagram of $\hat {L}_{pol}\frak g$ (see Table II
of [KW]) correspond exactly to $Out(K)\alpha C(K)$.  Given an element
of the center, we obtain an outer automorphism of $\hat {L}_{pol}
\frak g$,
and also of $L_{pol}\frak g$ (and its completions). There is a
canonical way to realize this outer automorphism by
permuting the $r+1$ copies of $sl(2,\Bbb C)$ which generate
$\hat {L}_{pol}\frak g$.  This automorphism preserves the
triangular decomposition of $\hat {L}_{pol}\frak g$, it commutes with the
unitary involution, and it also lifts to an
automorphism of $\hat {L}_{pol}G$ (and its completions).
At the level of loops, this
automorphism is realized by
a multivalued loop in
$\Cal L_{pol}Aut(K)$ (such as conjugation by $\bold w_0\left(\matrix
z^{1/2}&0\\
0&z^{-1/2}\endmatrix \right)$,
which we used in
$\S 2-3$); see $(3.4.2)$ and $(3.4.4)$ of [PS].
Asymptotic invariance of
Wiener measure implies that the limiting
distribution properties of $g\in \bold LG$, and in particular $a_
0$,
are invariant under this symmetry.  The formulas
we obtain below for this symmetry imply Lemma
$(5.14)$. This requires a case by
case analysis.

With $\nu_{\beta}$ probability one, $g=g_{-}\cdot g_0\cdot g_{+}$ is in the top
stratum of $\bold LG$, and in turn $g_0=l_0m_0a_0u_0$, where $l_0
\in N^{-}$,
$m_0\in T$, $a_0\in exp(\frak h_{\Bbb R})$, and $u_0\in N^{+}$.  We can write
$a_0=\prod_1^ra_j^{h_j}$, where the $h_j$ are the coroots of the simple
positive roots $\alpha_j$, and $r$ denotes the rank of $\frak g$.  In turn
we can write
$$a_j=\vert\sigma_j(\hat {g})/\sigma_0(\hat {g})^{\check {a}_j}\vert
=\vert\langle\pi_{\Lambda_j}(g_0)v_{\Lambda_j},v_{\Lambda_j}\rangle
\vert ,\tag B.1$$
where the $\sigma_j$ are the fundamental matrix coefficients,
viewed as functions on $\hat {\bold L}G$, $\hat {g}\in\hat {\bold L}
G$ projects to $g$, the
integers $\check {a}_j$ are determined by $h_{\theta}=\sum\check {
a}_jh_j$
(see Table I of [KW]), the $\Lambda_j$ are the fundamental
dominant weights for $\frak g=\frak n^{-}\oplus \frak h\oplus \frak n^{
+}$ corresponding to the
$\alpha_j$, and the $\pi_{\Lambda_j}$ denote the corresponding irreducible
representations (for the first equality in $(B.1)$, see \S 4.3
of Part III of [Pi1]).

The main point is that in all cases with $C(K)\ne 0$, there
is a corresponding automorphism which interchanges $\sigma_0$
and some other $\sigma_j$, $j>0$ (see Table II of [KW]; note that
automorphisms which fix the $0$ node correspond to outer
automorphisms of $K$). This is
what leads to the inversion formulae in Lemma $(5.14)$.

Suppose that $K=SU(n)$.  In this case $C(SU(n))=\Bbb Z_n\Delta$,
where $\Delta =exp(2\pi i/n)$.  The generator $\Delta$ corresponds to
the Dynkin diagram symmetry $\alpha_j\to\alpha_{j+1}$, where $j$ is read
mod $n-1$.  All $\check {a}_j=1$.  We want to write down the
multivalued loop which corresponds to the automorphism
of $\hat {L}\frak g$ which sends $h_j$, $e_j$, $f_j$ to $h_{j+1}$, $
e_{j+1}$, $f_{j+1}$, where $j$
is read mod $n-1$. In the defining representation of $sl(n,\Bbb C
)$,
where $\frak h$ is the diagonal subalgebra, this multivalued loop
is given by
$$\sigma_{\Delta}(t)=\lambda \bold w\prod_{j=1}^{n-1}e^{2\pi i(\frac
jn)th_j}$$
$$=\lambda\left(\matrix 0&0&..&0&exp(-2\pi i\frac {n-1}nt)\\
exp(2\pi it/n)&0&..\\
0&exp(2\pi it/n)&0\\
..\\
0&0&..&exp(2\pi it/n)&0\endmatrix \right)$$
where $\bold w$ is the matrix representing the cyclic
permutation $(12..n)$, and the factor $\lambda$ guarantees $det=1$,
and is otherwise irrelevant.  It is straightforward to
check that conjugation by $\sigma_{\Delta}$ implements the
automorphism described above.

The automorphism of the Kac-Moody extension cyclically
permutes $\sigma_0$, $\sigma_1$$,..,\sigma_{n-1}$.  Together with $
(B.1)$, this
implies that $\sigma_{\Delta}$ maps $(a_j)$ to $(a_2/a_1,a_3/a_1,
..a_{n-1}/a_1,a_1^{-1})$.
Thus in this case $a_1^{-1}$ has the same limiting distribution
properties as $a_{n-1}$.

In the case of $C_l$, $C(Sp(l))=\Bbb Z_2\Delta$, where $\Delta =-
1$ (as a
matrix in the defining representation).
This transposes $\sigma_0$ and $\sigma_l$. Hence $(a_j)$ maps to
$(a_1/a_l,..a_{l-1}/a_l,1/a_l)$. We have
$$\sigma_{\Delta}(t)=\bold w\prod_{j=1}^lexp(i\pi tjh_j)$$
where (if we realize $C_l$ as in [V], page 300) $\bold w$ is
a particular
representative for the Weyl group element that
maps $(\lambda_1,..,\lambda_l)$ to $(-\lambda_l,..,-\lambda_1)$ (so that $
h_i\leftrightarrow h_{l-i}$, $i<l$, and
$h_l\leftrightarrow -h_{\theta}$). The formula is
$$\bold w=\left(\matrix A&0\\
0&A^{-1}\endmatrix \right)\left(\matrix 0&B\\
-B&0\endmatrix \right)$$
where $A=diag(i,-i,i,..)$, $B$ has $1$'s on the antidiagonal,
and $A$ and $B$ are $l\times l$ matrices.

In the case of $B_l$, $C(Spin(2l+1))=\Bbb Z_2\Delta$, the kernel
of the projection to $SO(2l+1)$. The corresponding automorphism
transposes $\sigma_1$ and $\sigma_0$; hence $(a_j)$ maps to
$(1/a_1,a_2/a_1,..,a_l/a_1)$. We must have
$$\sigma_{\Delta}(t)=\bold wexp(i\pi th_l),$$
where $\bold w$ is a representative for the Weyl group element
that
changes the sign of $\lambda_1$ (this implies $h_1\leftrightarrow
-h_{\theta}$).

The other cases are similar.

\bigskip

\flushpar {\bf The Type I Case.  }

\smallskip

Suppose that there exists $\Delta\in C(U)$ which is fixed by $\Theta$.
In terms of the classification of type I symmetric
spaces, this applies in all cases except the family
$SU(n)/SO(n)$, for odd $n$ (because $SO(n)$ does not have a
center for odd $n$), $E_6/Sp(4)$ (the central element of
$Sp(4)$, of order $2$, is not contained in $C(E_6)\simeq \Bbb Z_3$), $
E_6/F_4$,
and all cases in which $U=G_2,F_4$ or $E_8$ (because these
do not have central elements).

Given $\Delta$ fixed by $\Theta$, the corresponding automorphism of
$\hat {L}\frak g$ will commute with the unique lift of $\Theta$ which fixes
$c$, the central element.  We therefore obtain a loop $\sigma$
which will act on $L(U/K)$ and our distributional
completions.  We write $g_0=l_0\bold wma_0l_0^{*\Theta}$.

In cases in which $\Theta$ is an inner automorphism,
$\frak h_0=\frak t_0=\frak t$, and the formulae for the action the $
a_j$ above
apply.

Suppose $U/K=SU(2n)/SO(2n)$, where $\Theta$ is realized via
the canonical (outer) automorphism which interchanges
$\alpha_1\leftrightarrow\alpha_{2n-1}$, $\alpha_2\leftrightarrow\alpha_{
2n-2}$, ...  We have $\Delta =-1$.  Then $a_0$ is of
the form
$$a_0=a_1^{h_1}..a_n^{h_n}a_{n-1}^{h_{n+1}}..a_1^{h_{2n-1}}$$
and the automorphism corresponding to $\Delta =-1$ (the $n^{th}$
power of the automorphism corresponding to $\sigma$ in $(B.2)$)
maps this to the sequence of $a_j's$
$$(a_2/a_n,..,a_{n-1}/a_n,a_1/a_n,1/a_n,a_1/a_n,a_{n-1}/a_n,..,a_
2/a_n)$$
Thus $1/a_n$ is equivalent to $a_{n-1}$ in distribution in the
limit $\beta\to 0$.

The case $U/K=SU(2n)/Sp(n)$ is similar.

\bigskip

\centerline{Appendix C. Integral Formulas for $G/G_{\Bbb R}$.}

\bigskip

Consider the Cartan decomposition
$$\psi :U\times_Ki\frak k\to G/G_{\Bbb R}:[g,x]\to ge^xG_{\Bbb R}\tag C.1$$
(here $K$ acts on $U$ on the right and on $i\frak k$ by the adjoint
action, and given $g\in U$, $x\in i\frak k$, $[g,x]$ denotes the image
point in $U\times_Ki\frak k$).  We want to express the $G$-invariant
measure on $G/G_{\Bbb R}$ (which is unique up to a scalar) in
terms of these coordinates.  To do this we recall that
there is a natural $U$-invariant connection on the vector
bundle
$$U\times_Ki\frak k\to U/K.\tag C.2$$
The horizontal subspace at the point $[g,x]\in U\times_Ki\frak k$ is
defined to be the image of the map
$$i\frak p\to Hor\vert_{[g,x]}:\zeta\to\frac d{dt}\vert_{t=0}[ge^{
t\zeta},x];\tag C.3$$
this map depends upon the choice of $(g,x)$, but the image
is independent of this choice.  Given this horizontal
distribution, we obtain a Riemannian structure on our
vector bundle, using the inner products on $i\frak p$ and $i\frak k$ (the
vertical direction) induced by the Killing form.  We let
$dV$ denote the corresponding volume element.

Let $dV_{G/G_{\Bbb R}}$ denote the $G$-invariant volume for $G/G_{
\Bbb R}$
(this exists because the adjoint action of $G_{\Bbb R}$ on $i\frak g_{
\Bbb R}$
admits an essentially unique invariant volume form).

\proclaim{Proposition (C.4)}We have
$$\psi^{*}(dV_{G/G_{\Bbb R}})=c\prod_{\alpha >0}^{\frak p}cosh^2(
\alpha (x_0))\prod_{\alpha'>0}^{\frak k}\vert\frac {sinh\alpha'(x_
0)}{\alpha'(x_0)}\vert^2dV([g,x])$$
where $x\in i\frak k$ is $K$-conjugate to $x_0\in i\frak t_0$, and the products
are over the positive roots for $i\frak t_0$ acting on $\frak p^{
\Bbb C}$ and $\frak k^{\Bbb C}$,
respectively.
\endproclaim

\flushpar Examples (C.5).  (a) If $X=S^2$, then
$$dV_{SL(2)/SU(1,1)}=cosh^2(2\vert x\vert )dV,\tag C.6$$

(b) In the group case $X=K$, there is a more direct
formulation of this result.  In this case there is a
commutative diagram
$$\matrix (K\times K)\times_{\Delta (K)}i\frak k&\to&(K^{\Bbb C}\times
K^{\Bbb C})/\{(g,g^{-*}):g\in K^{\Bbb C}\}\\
\downarrow&&\downarrow\\
K\times i\frak k&\to&K^{\Bbb C}\endmatrix \tag C.7$$
where the first vertical arrow is given by
$$(K\times K)\times_{\Delta (K)}i\frak k\to K\times i\frak k:[(g,
h),x]\to (gh^{-1},2hxh^{-1})\tag C.8$$
the second vertical arrow is given by $[g,h]\to gh^{*}$, and the
horizontal arrows are of the form $(g,x)\to ge^x$.  Note the
essential appearance of the ``$2$'' in $(C.8)$.  In terms of the
coordinates $g=ke^{x'}$, for $g\in K^{\Bbb C}$,
$$dg=c\prod_{\alpha >0}\vert\frac {sinh(\alpha (x'_0))}{\alpha (x'_
0)}\vert^2dk\times dx'\tag C.9$$
This is equivalent to $(C.4)$, because up to conjugation
$x'=2x$ and $sinh(2x)=2cosh(x)sinh(x)$, and the $\frak p$ and $\frak k$
roots are the same.

\smallskip

\demo{Proof of (C.4)} Fix $g\in U$ and $x\in i\frak k$.  Given these
choices, using the map $\psi$ in $(C.1)$, we can identify
$$T(U\times_Ki\frak k)\vert_{[g,x]}=Hor\oplus Vert=i\frak p\oplus
i\frak k=i\frak g_{\Bbb R}\tag C.10$$
We also have an identification
$$i\frak g_{\Bbb R}\to T(G/G_{\Bbb R})\vert_{G_{\Bbb R}}.\tag C.11$$
Thus given our choice of representative $ge^x$ for the
coset $ge^xG_{\Bbb R}$, we obtain an identification
$$T(G/G_{\Bbb R})\vert_{ge^xG_{\Bbb R}}=i\frak p\oplus i\frak k=i
\frak g_{\Bbb R},\tag C.12$$
where $X\in i\frak g_{\Bbb R}$ corresponds to
$$\frac d{dt}\vert_{t=0}ge^xe^{tX}\cdot G_{\Bbb R}.\tag C.13$$
The $G$-invariant volume form at $ge^xG_{\Bbb R}$ corresponds to
the $G_{\Bbb R}$-invariant volume form on $i\frak g_{\Bbb R}$ via this
identification.

With these identifications understood, we claim that
$$d\psi\vert_{[g,x]}:i\frak p\oplus i\frak k\to i\frak p\oplus i\frak k
:\zeta ,y\to proj_{i\frak g_{\Bbb R}}(e^{-adx}(\zeta )+\frac {1-e^{
-adx}}{adx}(y))).\tag C.14$$
To verify this we calculate
$$\frac d{dt}\vert_{t=0}e^{-x}g^{-1}ge^{t\zeta}e^{x+ty}=\frac d{d
t}\vert_{t=0}(e^{tAd(e^{-x})(\zeta )}e^{-x}e^{x+ty})\tag C.15$$
$$=e^{-adx}(\zeta )+\frac {1-e^{-adx}}{adx}(y).\tag C.16$$
When we project, we obtain the claim.

Now we observe that since $x\in i\frak k$, $ad(x)$ maps $i\frak g_{
\Bbb R}$ to $\frak g_{\Bbb R}$,
and $ad(x)^2$ maps $i\frak g_{\Bbb R}$ into itself.  Thus with respect to
the decomposition $i\frak p\oplus i\frak k$,
$$d\psi\vert_{[g,x]}=\left(\matrix cosh(adx)&0\\
0&\frac {sinh(adx)}{adx}\endmatrix \right).\tag C.17$$
We then have
$$det_{\Bbb R}(\frac {sinh(adx)}{adx}:i\frak k\to i\frak k)=det_{
\Bbb C}(\frac {sinh(adx)}{adx}:\frak k^C\to \frak k^{\Bbb C})=\prod_{
\alpha >0}^{\frak k}(\frac {sinh(\alpha (x_0))}{\alpha (x_0)})^2.\tag C.18$$
and similarly for $cosh(adx)$.  This proves $(C.4)$.  \qed
\enddemo

We now consider the triangular decomposition.

\proclaim{Proposition (C.19)}We write an element of the
top stratum of $\phi (G/G_{\Bbb R})$ as in $(d)$ of $(1.8)$,
$g=l\bold wma_{\phi}l^{*\Theta}$.  In these coordinates
$$dV=\bold a_{\phi}^{\rho}dm(\bold a_0)dm(l)dm(m)dm(\bold w),$$
where $\rho$ denotes the sum of the positive complex roots.

\endproclaim

\demo{Proof of (C.19)}$B^{-}$ has open orbits in the top
stratum.  The $G$-invariant measure in these orbits will
be determined by $B^{-}$-invariance.  The measure $dV$ is
clearly $B^{-}$-invariant.  \qed
\enddemo

\bigskip

\centerline{References}

\bigskip

\flushpar[B] P Billingsley, Convergence of Probability
Measures, John Wiley and Sons (1968).

\flushpar[D] P DeLorme, Harmonic analysis on real
reductive symmetric spaces, Proceedings of the ICM
2002, Beijing, Vol II, World Scientific Publishing (2003)
545-554.

\flushpar[H] S Helgason, Differential Geometry, Lie
Groups, and Symmetric Spaces, Academic Press (1978).

\flushpar[K] A Kocherlakota, Homology of real flag
manifolds and loop spaces of symmetric spaces, Adv
Math, Vol 110, Issue 1, (1995) 1-46.

\flushpar[KW] V Kac and M Wakimoto, Modular and
conformal invariance constraints in representation
theory of affine algebras, Adv Math 70 (1988) 156-237.

\flushpar[MM] M and P Malliavin, Integration on loop
groups III.  Asymptotic Peter-Weyl orthogonality, J
Funct Anal 108 (1992) 13-46.

\flushpar[Pi1] D Pickrell, Invariant measures for unitary
forms of Kac-Moody Lie groups, Memoirs of the AMS,
Vol 146, No.  693 (2000).

\flushpar[Pi2] --------, The radial partial of the
zero-mode Hamiltonian for the sigma model with group
target space, Rev Math Phys, Vol 16, no.  5 (June
2004), 603-628.

\flushpar[Pi3] --------, Examples of $PSU(1,1)$-invariant
probability measures on $H^m(\Delta )$, in preparation.

\flushpar[Pi4] ---------, The diagonal distribution of the
invariant measure of a compact symmetric space, in
preparation.

\flushpar[PS] A Pressley and G Segal, Loop Groups, Oxford
Mathematical Monographs, Oxford Press (1986).

\flushpar[V] V S Varadarajan, Lie Groups, Lie Algebras,
and Their Representations, Springer-Verlag (1984).

\end

$================================$

The other cases are similar.

In the case of $C_l$, $C(Sp(l))=\Bbb Z_2\Delta$, where $\Delta =-
1$ (as a
matrix in the defining representation).
This transposes $\sigma_0$ and $\sigma_l$. Hence $(a_j)$ maps to
$(a_1/a_l,..a_{l-1}/a_l,1/a_l)$.

For $D_l$, $l$ odd, $C(D_l)=\Bbb Z_4\Delta$,
$\Delta =exp(i\pi (h_1+h_3+..+h_{l-2}+\frac 12(h_{l-1}-h_l)))$. This

For $l$ even,
$C(D_l)=\Bbb Z_2exp(i\pi (h_{l-1}+h_l))\times \Bbb Z_2exp(i\pi (h_
1+h_3+..h_{l-1}))$.

 $a_1$
and $1/a_1$ are equivalent ($D_l$ has some additional
symmetry);

For $E_6$, $C(E_6)=$$\Bbb Z_3\Delta${\cyr ,} $\Delta =exp()${\cyr .}We can take $
\sigma_{\Delta}$ to be
. This maps $(a_j)$ to
$(a_5/a_1,a_4/a_1,a_3/a_1,a_6/a_1,1/a_1,a_2/a_1)$, hence $a_5$ and $
1/a_1$
are equivalent.  For $E_7$, $(a_j)$ maps to
$(a_5/a_6,a_4/a_6,a_3/a_6,a_2/a_6,a_1/a_6,1/a_6,a_7/a_6)$, hence  $
a_6$
and $1/a_6$ are equivalent.

$========================$

\centerline{\S 6. Loops in More General Spaces}

\bigskip

In this final section, to put this paper in perspective,
we discuss other kinds of target spaces.  The basic
theme is that the behavior of Wiener measures, for
large temperature, seems to reflect some aspects of
renormalization group analysis of the sigma model with
the given target space (see \S\ of [Ga]).  Whether these
things are tangibly related is not clear.

\bigskip

We can definitely generalize this paper in the direction
of classical infinite symmetric spaces and twisted loop
spaces.  All of this fits into the framework of
Kac-Moody theory.

Do we want to give a short rundown on what is known
about classical symmetric spaces.  Especially the fact
that the Laplacian is a fuzzy sort of operator globally,
but makes sense as a differential operator in radial
coordinates, which are not smooth.

\bigskip

\centerline{Compact Homogeneous Spaces}

\bigskip

Suppose that $\theta$ is an arbitrary automorphism of $U$ (a
simply connected compact Lie group).  As in the
symmetric case, we obtain groups
$$\matrix &&G\\
&\nearrow&&\nwarrow\\
H&&&&U\\
&\nwarrow&&\nearrow\\
&&K\endmatrix $$
where $H$ is the fixed point set of $(\cdot )^{-*\theta}$.  For example if
$\theta$ is a generic inner automorphism, i.e.  conjugation by a
regular element $h\in U$, then $K=T$, a maximal torus, and
$H$ is the corresponding Cartan subgroup of $G$.

We also still obtain mappings
$$\matrix U/K&\to&U\\
\downarrow&&\downarrow\\
G/H&\to&G\endmatrix :gH\to gg^{*\theta}.$$
and presumably the down arrow inclusions are still
homotopy equivalences.  Also $U/K$ is still simply
connected by Bott's theorem.  Note that in the generic
case $U/K$ is the conjugacy class of $h$ right shifted by
$h^{-1}$, hence has the topological type of the flag space.

However in the case in which $\theta^2\ne 1$, we do not
apparently have simple algebraic equations which cut out
these embeddings.

Barring the existence of these equations, this suggests
that when we compute the limit of Wiener measure on
$L(U/K)$, as temperature tends to infinity, we obtain the
measure $\delta_{LU}$.

\proclaim{Conjecture}When $\theta^2\ne 1$, Wiener measure for
$L(U/U^{\theta})$ coverges to the canonical measure associated to
$LU$ as temperature tends to $\infty$.
\endproclaim

This possibly has something to do with the geometric
fact that the holonomy for $U/K$ spreads out in a generic
way, unless $U/K$ is symmetric.  It would be interesting
to ask, in renormalization group analysis, does the sigma
model with target $U/U^{\theta}$, go to the sigma model for $U$.
[The idea here is that the Ricci flow tends have locally
symmetric limits, at least in low dimensions, so possibly
the sigma model for $U/U^{\theta}$ is in the same universality
class as $U$, at least using our type of embedding].

\bigskip

\centerline{General Simply Connected Compact $X$ }

\bigskip

Suppose now that $X$ is a general simply connected
compact Riemannian $X$.  One important process that we
definitely like to understand is how to go from $X$ to a
``zero-mode'' space.  For example given $X=K$, we obtain
$G$.  Given a symmetric $X=U/K$, we obtain $G/G_{\Bbb R}$.  In
the case of a compact homogeneous space corresponding
to a generic automorphism $\theta ,$ it appears that we simply
obtain $G$, and we lose contact with $X$.  At this point we
still do not know what the Riemannian geometry is on $G$
or $G/G_{\Bbb R}$ that corresponds to the zero-mode Hamiltonian.

Now suppose that $X$ is something like a Calabi-Yau
space.  We obtain a quantum sigma model.  Assuming the
space is not flat, there is a mass, or at least conformal
invariance is broken, but in the renormalization group
scheme, to first loop approximation, conformal symmetry
is intact.  This should be somehow reflected in what
happens to Wiener measure in the large temperature
limit.  For example suppose that $X=T$, a torus.  In this
case the zero-mode space is $T$ itself.  That is, the
measure does not spread out.  This is undoubtedly linked
to the absence of holonomy.

\bigskip

\centerline{Locally Symmetric Negatively Curved Spaces}

\bigskip

Let's return to the diagram
$$\matrix &&G\\
&\nearrow&&\nwarrow\\
G_{\Bbb R}&&&&U\\
&\nwarrow&&\nearrow\\
&&K&\endmatrix $$
From this we obtain a geodesic embedding
$$U/K\to G/G_{\Bbb R}=U\times_Ki\frak k$$
although the geometry on the target does not actually
seem relevant to anything connected with sigma models.

We now want to imitate this for the dual nonunitary
type symmetric spaces.  This essentially means that we
repeat our previous development, switching the roles of
$G_{\Bbb R}$ and $U$.  There are some interesting changes in the
results along the way.

We have natural maps
$$\matrix K&\to&G_{\Bbb R}&\to&G_{\Bbb R}/K\\
\downarrow&&\downarrow&&\downarrow\\
U&\to&G&\to&G/U\endmatrix .$$
In this context, in which the roles of $U$ and $G_{\Bbb R}$ have
been reversed, there are changes in the topology of
these maps.  For example the first two vertical
inclusions are no longer homotopy equivalences (in
general).  Only the last vertical arrow is a homotopy
equivalence, in which case both spaces are actually
contractible.  We presumably continue to have a
diffeomorphism
$$G_{\Bbb R}\times_{Ad(K)}i\frak k\to G/U,$$
given by the formula $(g,X)\to gexp(X)$ (where this is
interpreted as a coset representative), so that we can
continue to interpret $G/U$ as a kind of curvature or
holonomy bundle.

In turn there are totally geodesic embeddings (or
morphisms) of symmetric spaces
$$\matrix G_{\Bbb R}/K&\to&G_{\Bbb R}&:&gK&\to&gg^{-\theta}\\
\downarrow&&\downarrow\\
G/U&\to&G&:&gG_{\Bbb R}&\to&gg^{*}\endmatrix $$
(see \S\ of [Helg]).

From now on we will identify $G_{\Bbb R}/K$, $G_{\Bbb R}$, and $G
/U$ with
their images in $G$ with respect to these embeddings.
These images are defined by simple algebraic equations
(slightly more complicated than in the unitary case,
because there are some connectedness issues which
arise, and I do not know how to express this in general,
so it will temporarily be ignored)
$$\matrix G_{\Bbb R}/K\subset \{g\in G:g^{-1}=g^{*\theta}=g^{\theta}
\}&\to&G_{\Bbb R}=\{g^{-1}=g^{*\theta}\}\\
\downarrow&&\downarrow\\
G/U\subset \{g^{*}=g\}&\to&G\endmatrix $$

For example suppose we consider the $H^2$ case.  We have
$$\matrix &&SL(2,\Bbb C)\\
&\nearrow&&\nwarrow\\
SU(1,1)&&&&SU(2)\\
&\nwarrow&&\nearrow\\
&&U(1)\endmatrix $$
where
$$\left(\matrix a&b\\
c&d\endmatrix \right)^{\theta}=\left(\matrix a&-b\\
-c&d\endmatrix \right).$$
and $(\cdot )^{*}$ is the usual Hermitian conjugate.

Suppose that
$$g=g_{-}\cdot g_0\cdot g_{+}$$
is a point in the top stratum of $L_{hyp}(G/U)$.  This
implies that $g_0^{*}=g_0$, and $g_{-}=g_{+}^{*}$, i.e.
$$g_0=\left(\matrix a_0&b_0\\
\bar {b}_0&d_0\endmatrix \right),\quad a_0,d_0\in \Bbb R,\quad a_
0d_0-\vert b_0\vert^2=1,$$
Do we have $a_0>0$?  This may not be enforced.  If
$$g_{+}=\left(\matrix a(z)&b(z)\\
c(z)&d(z)\endmatrix \right)=1+\left(\matrix a_1&b_1\\
c_1&-a_1\endmatrix \right)z+..$$
then
$$g_{-}=\left(\matrix A(z)&B(z)\\
C(z)&D(z)\endmatrix \right)=g_{+}^{*}=1+\left(\matrix \B {a}_1&\bar {
c}_1\\
\bar {b}_1&-\bar {a}_1\endmatrix \right)z^{-1}+..$$

The action of the constants $G_{\Bbb R}$ on $L_{hyp}(G/U)$ is
completely transparent in these coordinates:  for $u\in G_{\Bbb R}$,
$$u\cdot g=[ug_{-}u^{-1}]\cdot [u^{}g_0u^{-\theta}]\cdot [u^{\theta}
g_{+}u^{-\theta}].$$

What do we have to augment the constants with to
generate $LG_{\Bbb R}$?  This action is not going to be of much
use, because Wiener measure with not be quasiinvariant.
We just have to study Wiener measure period.

\bigskip

Suppose that we consider the sigma model with target
$X=\pi\backslash G_{\Bbb R}/K$, where $\pi$ is a discrete cocompact subgroup,
e.g.  $G_{\Bbb R}=PSU(1,1)$ and $\pi$ is a surface subgroup.  The
loop space has a $\pi_0$ decomposition over $\pi$
$$LX=\bigsqcup_{\gamma\in\pi}Map_{\gamma}(\Bbb R,G_{\Bbb R}/K)$$
So even though the target is initially compact, we end
up considering maps into a noncompact target such as
$H^2$.

Now suppose that we put Wiener measure on this loop
space.  The question is what happens to this measure as
temperature tends to infinity.  We might be able to do
the Riemann-Hilbert factorization in a more explicit way
when we are considering positive symmetric matrices -
this is discussed in Gohberg-Krein.  And I think that
McKean has some explicit stuff related to Wiener
measure on positive matrices that might be useful.
Because we will be moding out by $\pi$, it is not out of the
realm of possibility that the zero-mode is something
interesting, for we have some compact directions,
corresponding to the projection
$$\pi\backslash G/U=\pi\backslash G_{\Bbb R}\times_Ki\frak k\to\pi
\backslash G_{\Bbb R}/K$$
In fact the space $\pi\backslash G_{\Bbb R}$ itself is compact, and this is
locally equivalent to this space.

------------------------------

For example in the case of a torus, our Wiener measures
simply push off to infinity - the probabilistic mass
escapes, except for the zero mode, where it simply
become Haar measure.  The vacuum appears in another
way.  There must be some conservation principle
involved here.  When the probabilistic mass goes to
infinity, something else appears in its place.  We need
to understand these conformal field theories involving $T$
much more thoroughly.

In the case of $T$ we end up reconstituing the vacuum by
introducing something like a level, although it is more
complicated, because we have to take into account the
inner product on $\frak t$, which is not sharply constrainted by
symmetry in this case.  However the bump is
apparently gotten by considering the determinant of a
Toeplitz operator.

In the case of $G_{\Bbb R}/K$, we can map this into $G/U$, and
we apparently have a global coordinate
$$L(G_{\Bbb R}/K)\to L(G/U)=G/U\times H^1(\Delta ,\frak g)$$
How do we put together the vacuum in this case?  How
do we interpret mass, etc.  This seems a great deal like
the Virasoro case, where the topology disappears, but
somehow there is underlying algebra that makes nice
things happen.  In this context we should keep in mind
Kac's results on unitary for representations of LSU(1,n).
Maybe this has some relevance, or possibly the results
of Zuckermann on coset type constructions for more
general noncompact type semisimple groups and their
loop spaces.

See Gawedzki for how one formally evaluates functional
integrals for the sigma model with target $G_{\Bbb R}/K$.  There
is a close connection here with the inner product for
Chern-Simons states, Yang-Mills in two dimensions, and a
host of other things.

We could go further and consider a Minkowskian type
target.  This gives us some nonflat examples which
would be interesting to consider, beyond the flat case
that produces the noghost theorem (we need to go
through this noghost construction, following Segal's
approach; this would be really enlightening).

There are dualities that we could consider in nonflat
situations.  But now we seem to need a more abstract
approach ala Segal.

\bigskip

\centerline{Geometry}

\bigskip

Consider the diagram of geodesic embeddings
$$\matrix U/K&\to&U\\
\downarrow&&\downarrow\\
G/G_{\Bbb R}&\to&G&\leftarrow&G/U\\
&&\uparrow&&\uparrow\\
&&G_{\Bbb R}&\leftarrow&G_{\Bbb R}/K\endmatrix $$
When we do Riemann-Hilbert factorization, we use the
complex structure of $G$.  Thus in looking for the
principal symbol for the zero-mode Hamiltonian, it is
natural to take this complex structure into account.

The most natural geometries for $G$ itself seem to either
the Riemannian geometry for $G$ gotten by identifying it
with $TK$ (this does not depend upon whether we use left
or right translation to identify $TK=K\times \frak k$, and the
Laplace operator is simply the product of the Laplacians
for $K$ and the linear space $\frak k$), and the Kahler geometry
for $G$ gotten from the Guillemin-Stenzel procedure.

Thus for $G/G_{\Bbb R}$ there are at least two natural
possibilities.  One is the Riemannian geometry coming
from the identification with $U\times_Ki\frak k$, and the other is the
Riemannian geometry induced by the Kahler geometry for
$G$.  The former is intrinsic, while the latter appears to
be extrinsic.  It is plausible that we are seeking
something extrinsic, in particular something that depends
upon the complex structure of $G$, because this plays a
key role in our construction through Riemann-Hilbert
factorization.

-------------------------------

Let's speculate about the latter geometry.  $G$ is Kahler,
and $G/G_{\Bbb R}$ is half the dimension of $G$.  This suggests
that $G/G_{\Bbb R}$ might be Lagrangian.  In fact there is the
possibility that there is some connection here with the
work of Krotz and Otto.  What is special about the
Riemannian structure for a Lagrangian?

\proclaim{Conjecture}$G/G_{\Bbb R}$ is a Lagrangian submanifold
of $G$
\endproclaim

\flushpar Proof.  In this calculation we will always use
the coordinates
$$U\times i\frak u\to G:g,x\to ge^x.$$
In these coordinates the symplectic structure is constant
and given by
$$\omega (\left(\matrix \xi\\
\eta\endmatrix \right),\left(\matrix \xi'\\
\eta'\endmatrix \right))=\langle i\xi\otimes\eta'-\eta\otimes i\xi'
\rangle$$
In the coordinates $()$, $G/G_{\Bbb R}$ is identified with
$$\{(g,x):g\in U/K\quad (i.e.g^{-1}=g^{\theta}),gx^{*}g^{-1}=x^{\theta}
\}.$$
The tangent space is identified with

One could ask the more basic question of whether the
map $G\to G:g\to g^{*\theta}$ is symplectic or antisymplectic.  It is
antiholomorphic.  If it were also something symplectic,
it would seemingly have some kind of isometry
properties.  This is possibly false.  So i question this
conjecture.

-------------------------------

\smallskip

We need to systematically study the geometry of $G/G_{\Bbb R}$.
What for example can we say about its ricci curvature.
Is $U/K$ geodesically embedded.  What can we say about
the geometry of the complementary directions.

This should be compared with string theory, which
starts with a space of the form $\Bbb R^{1,3}\times M$, where say $
M$
is Calabi-Yau.  I wonder if we can produce a space with
properties similar to this, using our construction.

\smallskip

Now suppose that we consider the noncompact side.  In
this case it is not so clear how the complex geometry of
$G$ can be related to some geometric structure for $G_{\Bbb R}$.

Inside of $G$ there is the $K^{\Bbb C}$-invariant covering of the
complex crown, the natural complexification of $G_{\Bbb R}$, that
Krotz talked about.  But I do not know how we get a
Laplacian out of that, related to $G/U$.  Maybe $G/U$ is
Lagrangian and positive, and we obtain a Riemannian
structure in this way.  All of this needs to be invariant
under discrete subgroups, so that we have some
compactness to work with.  We need the notion of a
positive Lagrangian.

\bigskip

\centerline{Vacuum Einstein Manifolds}

\bigskip

\bigskip

\bigskip

$==========================$

\flushpar\ (1.15) Basic Example.  If $X=S^n$, then we have
$$\matrix &&G=Spin(n+1,\Bbb C)\\
&\nearrow&&\nwarrow\\
G_{\Bbb R}=Spin(1,n,\Bbb R)&&&&U=Spin(n+1,\Bbb R)\\
&\nwarrow&&\nearrow\\
&&K=Spin(n,\Bbb R)\endmatrix ,\tag 1.16$$
where at the level of matrices, $\Theta$ is conjugation by
$diag(-1,1,1,..,1)$, and $A^{*}$ is the Hermitian conjugate of $A$.

In the special cases $n=2,3,4$ this is equivalent to
$$\matrix &&G=SL(2,\Bbb C)\\
&\nearrow&&\nwarrow\\
G_{\Bbb R}=SU(1,1)&&&&U=SU(2,\Bbb C)\\
&\nwarrow&&\nearrow\\
&&K=U(1)\endmatrix ,\tag 1.17$$
(when we identify $S^2$ with $\Bbb C\Bbb P^1$),
$$\matrix &&G=SL(2,\Bbb C)\times SL(2,\Bbb C)\\
&\nearrow&&\nwarrow\\
G_{\Bbb R}=SL(2,\Bbb C)&&&&U=SU(2)\times SU(2)\\
&\nwarrow&&\nearrow\\
&&K=SU(2)\endmatrix \tag 1.18$$
(when we identify $S^3$ with $SU(2)$, $K$ is embedded
diagonally, and $SL(2,\Bbb C)$ is embedded in $G$ as $g\to (g,g^{
-*})$),
and
$$\matrix &{{}\atop {}}&G=Sp(2,\Bbb C)\\
&\nearrow&&\nwarrow\\
G_{\Bbb R}=Sp(1,1)&&&&U=Sp(2)\\
&\nwarrow&&\nearrow\\
&&K=Sp(1)\times Sp(1)\endmatrix \tag 1.19$$
(when we are identifying $S^4$ with $\Bbb H\Bbb P^1$), respectively.
In this last case the $4$ dimensional space $S^4$ is
geodesically embedded into the $10$ dimensional, rank $2$
space $Sp(2,\Bbb C)/Sp(1,1)$.  This is an intriguing coincidence
of dimensions.

\flushpar Examples.  1)
$$\matrix &&SL(n,\Bbb C)\\
&\nearrow&&\nwarrow&\\
SL(n,\Bbb R)&&&&SU(n,\Bbb C)\\
&\nwarrow&&\nearrow\\
&&SO(n,\Bbb R)\endmatrix $$
where $\theta (g)=g^{-t}$, $\tau (g)=\bar {g}$.  We have
$$\phi :SL(n,\Bbb C)/SL(n,\Bbb R)\to SL(n,\Bbb C):g\to g\bar {g}^{
-1}$$
The image is $\{g\in G:g^{-1}=\bar {g}\}$.

Representatives for the $\theta$-stable Cartan subalgebras are
listed on page 129 of [Knapp].  We are interested in
$$\frak t_0+a_0=\{\left(\matrix \matrix t_1&\theta_1\\
-\theta_1&t_1\endmatrix &\\
&.\\
&&\matrix t_j&\theta_j\\
-\theta_j&t_j\endmatrix \\
&&&t_{[n/2]+1}\endmatrix \right):tr=0\}$$
where $j=[n/2]$, the $\theta_i$ correspond to $\frak t_0$, the $t_
j$
correspond to $\frak a_0$, and $t_{[n/2]+1}$ is only present if $
n$ is
odd.  [At the opposite extreme we can take $\frak h_0=$ $\frak a$, the
set of real diagonal matrices in $sl(n,\Bbb R)$, while $\frak t=0$; this
is because $G_{\Bbb R}$ is a normal real form.]

How do we identify the Weyl group with permutations,
from this point of view, so that we can understand how
$\Theta$ acts?  And we also want to understand how $w_0$ acts.
We need to find the simple roots $\alpha_1$$,..,\alpha_{n-1}$.  \smallskip

2)
$$\matrix &&Sp(n,\Bbb C)\\
&\nearrow&&\nwarrow\\
Sp(n,\Bbb R)&&&&USp(n)\\
&\nwarrow&&\nearrow\\
&&U(n)\endmatrix $$
In this case $\frak k$ has the same rank as everything else.
Hence we can forget about $\frak a$ and just take a maximal
torus in $U(n)$.  But this is also a normal real form case
as well!  Note that the symmetric spaces involved are
Hermitian.

We polarize $\Bbb C^{2n}$ in the usual way.  We have
$$\Theta (\left(\matrix A&B\\
C&D\endmatrix \right)=\left(\matrix A&-B\\
-C&D\endmatrix \right)$$
at both the group and Lie algebra level.  At the Lie
algebra level $B$ and $C$ must be symmetric in the
appropriate sense.

We have $\frak h_0=\frak t_0$, so we can divide roots into compact
and noncompact type.  The former are of the form
$\lambda_i-\lambda_j$ and the latter are of the form $\lambda_i+\lambda_
j$.  $\Theta$ acts
trivially on the $i_{\alpha}$ when $\alpha$ is compact and twists it in
the noncompact case.

\proclaim{Lemma} Suppose that $\left(\matrix a&b\\
c&d\endmatrix \right)\in SL(2,\Bbb C)$ and
$g=g_{-}\cdot g_0\cdot g_{+}$ is a point in the top stratum of $\bold L
(G/G_{\Bbb R})$,
so that $g^{*\theta}_0=g_0$ and $g_{+}=g_{-}^{*\theta}$, as in $(
)$.  We write $g_{\pm}$ as
in $()$, and $x_n=x_n'+Z_ne_{\theta}$, so that $y_n=y_n'+W_ne_{-\theta}$, where
$y_n'=x_n^{\prime *\theta}$, $W_n=-\bar {Z}_n$.  Let $h=i_0(\left
(\matrix a&b\\
c&d\endmatrix \right))$.  If $a+bZ_1\ne 0$,
then the point
$$g'=h\cdot \{g_{-}\cdot g_0\cdot g_{+}\}\cdot h^{*\theta}$$
is also in the top stratum, and has triangular
factorization $g'=g_{-}'\cdot g_0'\cdot g_{+}'$, where
$$g_{-}'=hg_{-}exp(-\frac b{a+bZ_1}e_{-\theta}z)g_0exp(\frac {-\bar {
b}}{\bar {a}-\bar {b}W_1'}e_{\theta}z^{-1})g_0^{-1}(a+bZ_1)^{h_{\theta}}
l_0^{-1},$$
$$g_0'=l_0(a+bZ_1)^{-h_{\theta}}g_0(\bar {a}-\bar {b}W_1')^{h_{\theta}}
u'_0,$$
and
$$l_0=exp(-ab[e_{-\theta},a^{-ad(h_{\theta})}(x_1')]-b^2Z_2e_{-\theta}
)\in N^{-}.$$
[If one calculates the left right action in the opposite
order, one obtains a second expression, not a priori
equivalent to the first,
$$g_{-}'=hg_{-}g_0exp(\frac {\bar {b}}{\bar {a}-\bar {b}W_1}e_{\theta}
z^{-1})g_0^{-1}exp(\frac b{a+bZ_1'}e_{-\theta}z)(a+bZ_1')^{h_{\theta}}
l_0^{\prime -1},$$
$$g_0'=l_0'(a+bZ_1')^{h_{\theta}}g_0(\bar {a}+\bar {b}\bar {Z}_1)^{
-h_{\theta}}u_0,$$
$$Z_1'=Z_1+\frac {\bar {b}}{\bar {a}+\bar {b}W_1}diag(g_0)^{\theta}$$
$$a+bZ_1'=\frac {\vert a+bZ_1\vert^2+b\bar {b}diag(g_0)^{\theta}}{
\bar {a}+\bar {b}\bar {Z}_1}$$
$$u_0=exp(-ac[e_{\theta},a^{-ad(h_{\theta})}(y_1')]-c^2W_2e_{\theta}
)\in N^{+}].$$

\endproclaim

$=========================$

\proclaim{Proposition} For $g$ in the top stratum of
$U/K\hookrightarrow G/G_{\Bbb R}$, as in $()$, we write
$$g=l\bold w_1m_1a_1G_{\Bbb R},$$
where $l\in N^{-}$, $\bold w_1\in N_U(T)/N_K(T)$, and $a_{\phi}\in
exp(i\frak t_0)$.  Then
$$\int_{U/K}a_{\phi}(g)^{-i\lambda}=\frac 1{\vert\pi_0\vert}\sum_{
\bold w_1}\prod^{\bold w_1}\frac {\langle\rho ,\alpha\rangle}{\langle
\rho -i\lambda ,\alpha\rangle}$$
where the $\bold w_1$ product is over the positive roots which
are imaginary and noncompact with respect to $Ad(\bold w)\Theta$.
\endproclaim

\flushpar {\bf Elementary Properties }

\smallskip

\proclaim{Proposition (5.24)}Suppose that $\mu$ is a
$L_{pol}K$-invariant probability measure on $\bold LG$.  Then $\mu$ is
supported on $L_{hyp}G$.
\endproclaim

The content of this proposition is that $g_{\pm}$ are genuine
holomorphic functions in $\Delta$ and $\Delta^{*}$, respectively.  If
Conjecture $(3.31)$ is true, then $g_{\pm}$ just barely miss
having radial limits along $S^1$ (this can possibly be
understood directly in terms of Wiener measure and
critical exponents; see $\S 2$ of [Pi2]).

\demo{Proof of (5.24)} We will work with the
representation $\pi_{\lambda}$ above.

By $(5.14)$ we know that $B_n'$ has
the standard invariant distribution.  This implies that
$\mu \{\vert B_n'\vert >R\}=(1+R^2)^{-1}$.  For $n=1$, $B_n'=\langle
\theta_1e_{-\theta},e_{\theta}\rangle$. is the
$e_{\theta}$ component of $\theta_1$. The
group $K$ acts on $\{\theta_1\}$ by its irreducible adjoint
representation. This implies that we can find $d$ elements
$g_i\in K$ such that the matrix coefficients
$\langle\theta_1e_{-\theta},Ad(g)(e_{\theta})\rangle$ span the dual. This implies
$$\{\vert\theta_1\vert >R\}\subset\bigcup_i\{\vert\langle\theta_1
,Ad(g)e_{\theta}\rangle\vert >R/d\},$$
hence
$$\mu \{\vert\theta_1\vert >R\}\le d/(1+(R/d)^2)^{-1}$$
where $d$ is the dimension of $\frak g$.  This gives us an
estimate for $A_1$.  Together with our estimate for $B_2'$, we
obtain an estimate for $B_2$.  This gives us an estimate
for $g_2$, and we continue.

We need to show that $\limsup_n\vert\theta_n\vert^{1/n}>1+\delta$  has measure
zero. So we need estimates for the probability that
$\vert\theta_n\vert >(1+\delta )^n$, at least for large $n$.

 \qed
\enddemo

\proclaim{Corollary}A measure $\mu$ as in $(5.24)$, viewed as
a measure on $L_{hyp}G$, is necessarily $L_{an}K$-invariant.
\endproclaim

\demo{Proof of ()}Given $g\in L_{an}K$, we can find $g_j\in L_{po
l}K$
such that $g_j$ converges to $g$ in the Lie group topology
for analytic loops described locally in the introduction
(see (3.5.3) of [PS]).  Given a Borel subset $E\subset L_{hyp}G$, the
characteristic functions for the sets $g_j(E)$ will converge
pointwise to $g(E)$.  Thus
$$\mu (g(E))=\lim_j\mu (g_j(E))=\lim_j\mu (E)=\mu (E).$$
\qed
\enddemo

Conjecturally the full symmetry group of $\mu$ is
$\Cal D_{W^{3/2}}\alpha \Cal L_{W^{1/2}}Aut(K)$, analogous to the fact that the full
symmetry group of $\nu_{\beta}$ is $Rot(S^1)\alpha \Cal L_{W^1}Au
t(K)$.

The involution which fixes $LK$ inside $LG$ does not extend
to our distributional completions.  However, at least in a
measure-theoretic sense, the support of $\mu$ (assuming
uniqueness) singles out a ``unitary form''.  Describing
this support in a precise way (analogous to Levy's law
for Brownian motion) is an attractive problem.  One
feature such a description should capture is the
following

\proclaim{Proposition}Suppose that $\mu$ is a $L_{pol}K$-invariant
probability measure on $\bold LG$.  Suppose further that $\mu$ is a
limit point of the Wiener measures $\nu_{\beta}$.  Then for each
$e^{i\theta}\in S^1$, as $r\uparrow 1$, the distribution of
$$g_{-}(r^{-1}e^{i\theta})g_0g_{+}(re^{i\theta})$$
converges weakly to $\delta_K$, the Haar distribution on $K$,
relative to $BC(G)$.
\endproclaim

This is true for each $\nu_{\beta}$, hence it is true for any limit
point, by a simple interchange of limits.

This proposition gives some meaning to the assertion
that $\mu$ is a kind of lift of the trivial invariant measure
$\prod_{S^1}\delta_K$.

\bigskip